\documentclass[
showpacs, showkeys,
 amsmath,amssymb,
 aps,
]{revtex4-1}

\pdfoutput=1
\usepackage{graphicx}
\usepackage{dcolumn}
\usepackage{bm}

\usepackage{subfigure}
\usepackage{longtable}
\usepackage{lipsum}
\usepackage{color}
\usepackage{comment}

\newcommand{\HeatFlux}{\boldsymbol{\mathcal{Q}}}
\newcommand{\SpeciesFlux}{\boldsymbol{\mathcal{F}}}
\newcommand{\SpeciesFluxB}{\bar{\boldsymbol{\mathcal{F}}}}
\newcommand{\StressTensor}{\boldsymbol{\Pi}}
\newcommand{\ShearViscosity}{\eta}
\newcommand{\BulkViscosity}{\kappa}

\newcommand{\EntropyProduction}{\mathfrak{v}}
\newcommand{\OnsagerMatrix}{\boldsymbol{\mathfrak{L}}}
\newcommand{\OnsagerMatrixB}{\bar{\boldsymbol{\mathfrak{L}}}}
\newcommand{\OnesVector}{\mathbf{u}}

\newcommand{\Jbar}{\bar{\mathbf{J}}}
\newcommand{\Xbar}{\bar{\mathbf{X}}}
\newcommand{\Lbar}{\bar{\mathbf{L}}}
\newcommand{\Bbar}{\bar{\mathbf{B}}}
\newcommand{\Wbar}{\bar{\mathcal{Z}}}

\newcommand{\Bcbar}{\bar{\mathcal{B}}}
\newcommand{\lbar}{\bar{\mathbf{l}}}
\newcommand{\xibar}{\bar{\xi}}
\newcommand{\DonevDiffusion}{{D}}
\newcommand{\U}{\mathbf{U}}
\newcommand{\FH}{\mathbf{F}_H}
\newcommand{\FD}{\mathbf{F}_D}
\newcommand{\FS}{\mathbf{F}_S}
\newcommand{\HH}{\mathbf{H}}
\newcommand{\Detoc}{\mathbf{D}^{f-c}}
\newcommand{\Dntoc}{\mathbf{D}^{n-c}}
\newcommand{\Gctoe}{\mathbf{G}^{c-f}}
\newcommand{\Gcton}{\mathbf{G}^{c-n}}
\newcommand{\GctoeT}{(\mathbf{G}^{c-f})^T}
\newcommand{\GctonT}{(\mathbf{G}^{c-n})^T}
\newcommand{\half}{\frac{1}{2}}

\global\long\def\V#1{\boldsymbol{#1}}
\global\long\def\M#1{\boldsymbol{#1}}

\global\long\def\d#1{\delta#1}

\global\long\def\grad{\M{\nabla}}

\global\long\def\av#1{\left\langle #1\right\rangle }

\begin{document}

\title{Fluctuating hydrodynamics of multi-species, non-reactive mixtures}

\author{Kaushik Balakrishnan,$^{1*}$ Alejandro L. Garcia,$^{2}$ Aleksandar Donev,$^{3}$ and John B. Bell$^{1}$}
\affiliation{$^1$ Computational Research Division, Lawrence Berkeley National Laboratory \\
 1 Cyclotron Road, Berkeley, CA 94720 \\ }
\affiliation{$^2$ Department of Physics and Astronomy, San Jose State University \\
 1 Washington Square, San Jose, CA 95192 \\ }
\affiliation{$^3$ Courant Institute of Mathematical Sciences, New York University \\
 251 Mercer Street, New York, NY 10012 \\ }

\date{\today}

\begin{abstract}

In this paper we discuss the formulation of the fluctuating Navier-Stokes (FNS) equations for multi-species, non-reactive fluids. In particular, we establish a form suitable for numerical solution of the resulting stochastic partial differential equations. An accurate and efficient numerical scheme, based on our previous methods for single species and binary mixtures, is presented and tested at equilibrium as well as for a variety of non-equilibrium problems. These include the study of giant nonequilibrium concentration fluctuations in a ternary mixture in the presence of a diffusion barrier, the triggering of a Rayleigh-Taylor instability by diffusion in a four-species mixture, as well as reverse diffusion in a ternary mixture. Good agreement with theory and experiment demonstrates that the  formulation is robust and can serve as a useful tool in the study of thermal fluctuations for multi-species fluids. The extension to include chemical reactions will be treated in a sequel paper.

\end{abstract}


\maketitle

\section{Introduction}
\label{sec:intro}

Since the pioneering work of Einstein and Smoluchowski on Brownian motion it has been clear that
hydrodynamic fluctuations are essential in the study of fluid dynamics at mesoscopic scales.
In fact, fluctuations play an important role in many physical, chemical, and biological processes,
ranging from phase separation to ion transport in cells.
For example, high-fidelity molecular simulations reveal that
thermal fluctuations significantly affect fluid mixing,
both in simple diffusion~\cite{Donev_11,DiffusionRenormalization}
and in the Rayleigh-Taylor instability \cite{Kadau_04,Kadau_07}.
The accurate modeling of droplets in nanojets \cite{Moseler_00,Eggers_02}
and lipid bilayer membranes~\cite{Atzberger2013A,Atzberger2013B}
necessitate the inclusion of hydrodynamic fluctuations.
Chemical processes, including combustion and explosive detonation, also depend
strongly on spontaneous thermal fluctuations~\cite{Nowakowski_03,Lemarchand_04}.
Finally, the manifestation of hydrodynamic fluctuations is not restricted to
mesoscale phenomena. Laboratory experiments involving gases, liquids or crystals demonstrate that,
away from equilibrium, thermal fluctuations lead to large-scale structures, the so-called
``giant fluctuation'' effect \cite{Vailati_97,Vailati_98,Vailati_11,Yuk_12}.

As an extension of conventional hydrodynamic theory, fluctuating hydrodynamics
incorporates spontaneous thermal fluctuations in a fluid
by adding stochastic flux terms to the deterministic fluid equations \cite{Zarate_07}.
These noise terms are white in space and time and are
formulated using fluctuation-dissipation relations
to yield the equilibrium covariances of the fluctuations.
This construction was first introduced by Landau and Lifshitz \cite{Landau_59} for a single component fluid.
Fox and Uhlenbeck \cite{Fox_70,Fox_70a} provide theoretical derivations of the fluctuation terms from perspectives of Brownian motion and the Boltzmann equation.
Numerous extensions of the theory have been developed, such as
to Extended Thermodynamics~\cite{Ikoma20112601} and plasma dynamics~\cite{PhysRevE.83.015401}.

The generalization of fluctuating hydrodynamics to binary mixtures was
first presented by Cohen \textit{et al.}~\cite{Cohen_71}
and by Law and Nieuwoudt~\cite{Law_89,Nieuwoudt_90}.
Multicomponent gaseous systems are discussed within the GENERIC framework
in the work of Ottinger~\cite{Ottinger_09}.
The standard fluctuating hydrodynamics theory for (thermo)diffusion in binary mixtures
(see, for example, Ortiz de Zarate and Senger \cite{Zarate_07}) has recently been
extended to non-ideal ternary mixtures in thermodynamic equilibrium by Ortiz de Zarate {\it et al.} \cite{TernaryEquilibriumFluct}.

Early work on numerical methods for the linearized fluctuating Navier-Stokes equations
was performed by Garcia {\it et al.} \cite{Garcia_87,Garcia_87a}.
More recently, we developed accurate and robust numerical techniques
for the full nonlinear system of equations~\cite{Bell_07,Bell_10,Donev_10,Balboa2012}.
In this paper, we extend to multicomponent systems the algorithm
developed for binary gas mixtures by Bell {\it et al.}~\cite{Bell_10} and
subsequently improved by Donev {\it et al.}~\cite{Donev_10}.

There are three reasons why this extension multi-species fluids is significant:
First, it allows us to consider interesting, realistic chemical reactions,
which will be the treated in a subsequent paper. Second, the majority of
microscopic systems of interest (and certainly all biological systems)
have negligible gradients of velocity and temperature.
The dominant mechanism for non-equilibrium entropy production in these systems
is from gradients of chemical potential (i.e., concentration gradients).
Third, there are interesting interaction effects due to
coupling of diffusion among the species.
In a single species fluid, the (deterministic) thermodynamic fluxes
are always in the direction of their conjugate thermodynamic force
(e.g., heat flux is always from hot to cold).
For a binary mixture there is an interaction between concentration and temperature
(e.g., heat flux due to a concentration gradient); however, this coupling is typically weak.
As we show in two examples in Section~\ref{sec:results},
diffusion barriers (zero concentration flux in the presence of a concentration gradient)
and reverse diffusion (concentration flux from low to high concentration)
can occur in multi-species mixtures (see for instance Duncan and Toor~\cite{DuncanToor}).
Giant fluctuations in binary mixtures out of
thermodynamic equilibrium have been studied for a long time, and here we demonstrate that the
coupling between the diffusive fluxes for different species also induces long-ranged correlations
between the concentrations of {\em different} species.

The paper is organized as follows:
The mathematical formulation is summarized in Section \ref{sec:form},
and the numerical scheme in Section \ref{sec:numerical}.
Computational results validating the methodology are presented in
Section \ref{sec:results} along with examples illustrating its capabilities.
Conclusions and directions for future work are discussed in Section \ref{sec:conclusions}.


\section{Theory}
\label{sec:form}

In this section, we summarize the mathematical formulation of the full multi-component, fluctuating Navier-Stokes (FNS) equations and establish the elements needed to develop a suitable numerical method for the resulting stochastic partial differential equations.
Our formulation of species diffusion is based on classical treatments,
such as in \cite{DM_63, BirdStewartLightfoot_06, NewIrrevThermoBook}.  We want to be able to utilize
existing software for computing transport properties of realistic gases such as the EGLIB  package \cite{EGLIB}
commonly used in the reacting flow community.  Consequently,  we will adopt the notation
given by Giovangigli \cite{Giovangigli_99}.
The formulation is general with the specific case of ideal gas mixtures treated in Section~\ref{IdealGasMixturesSection}.

\subsection{Multicomponent Hydrodynamic Equations}

The species density, momentum and energy equations of hydrodynamics are given by
\begin{equation}
\frac{\partial }{\partial t} \left( \rho_k \right)  + {\bf \nabla} \cdot \left( \rho_k {\bf v} \right) +
{\bf \nabla} \cdot {\SpeciesFlux}_k =  0, 
\label{eqn:spec}
\end{equation}
\begin{equation}
\frac{\partial }{\partial t} \left( \rho {\bf v} \right)
+ {\bf \nabla} \cdot \left[ \rho {\bf v} {\bf v}^T +  p \mathbf{I} \right]
+ {\bf \nabla} \cdot  {\StressTensor}  = \rho {\bf g},
\label{eqn:mom}
\end{equation}
\begin{equation}
\frac{\partial }{\partial t} \left( \rho E \right)  + {\bf \nabla} \cdot \left[ (\rho E + p) {\bf v}  \right] +
{\bf \nabla} \cdot \left[ {\HeatFlux} + \StressTensor \cdot {\bf v} \right] = \rho {\bf v \cdot g},
\label{eqn:energy}
\end{equation}
where $\rho_k$, ${\bf v}$, $p$, ${\bf g}$ and $E$ denote, respectively, the mass density for species $k$,
fluid velocity, pressure, gravitational acceleration, and total specific energy for a mixture with $N_s$ species
($k=1,\ldots N_s$).
Note that $\mathbf{v}\mathbf{v}^T$ is a (tensor) outer product with $T$ indicating transpose and
$\mathbf{I}$ is the identity tensor (i.e., ${\bf \nabla} \cdot p \mathbf{I} = {\bf \nabla} p$).
Transport properties are given in terms of
the species diffusion flux, ${\SpeciesFlux}$, viscous tensor, ${\StressTensor}$, and heat flux, ${\HeatFlux}$.
For Newtonian fluids, the deterministic viscous tensor is,
%
%
\begin{equation}
{\StressTensor} =
-\ShearViscosity \left ( \nabla \mathbf{v} + (\nabla \mathbf{v})^T  \right )
-\left(\BulkViscosity - \frac{2}{3} \ShearViscosity \right) \mathbf{I} \left({\bf \nabla} \cdot {\bf v}\right),
\end{equation}
where $\ShearViscosity$ and $\BulkViscosity$ are the shear and bulk viscosity, respectively.

Exact mass conservation requires that the species diffusion flux satisfies the constraint,
\begin{equation}
\sum_{k=1}^{N_s} \SpeciesFlux_k = 0,
\label{eq:det_sum_cond}
\end{equation}
so that summing the species equations gives the continuity equation.
\begin{equation}
\frac{\partial }{\partial t}  \rho   + {\bf \nabla} \cdot \left( \rho {\bf v} \right) = 0,
\label{eqn:cont}
\end{equation}
where the total density $\rho = \sum_{k=1}^{N_s} \rho_k$.
The mass fraction of the $k$-th species is denoted by $Y_k = \rho_k/\rho$ with $\sum_{k=1}^{N_s} Y_k = 1$.

In fluctuating hydrodynamics, we augment the fluxes in (\ref{eqn:spec})-(\ref{eqn:energy})
by adding a zero-mean stochastic flux to the deterministic flux.
For example, the viscous tensor becomes
${\StressTensor} + \widetilde{\StressTensor}$
where $\langle \widetilde{\StressTensor} \rangle = 0$
with $\langle \, \rangle$ denoting a suitably defined ensemble average.
The stochastic viscous flux tensor is a Gaussian random field that can be
written as~\cite{Landau_59,Espanol1998}
\begin{equation}
\widetilde{\StressTensor}(\mathbf{r},t) =
\sqrt{2 k_B T \ShearViscosity}\;
\widetilde{\mathcal{Z}}^{v}
+ \left ( \sqrt{\frac{k_B \kappa T}{3}} - \sqrt{\frac{2 k_B \eta T}{3}} \right ) \text{Tr} ( \widetilde{\mathcal{Z}}^{v}  ),
\label{StressTensorCorrelationEqn}
\end{equation}
where $k_B$ is Boltzmann's constant, $T$ is temperature and
$\widetilde{\mathcal{Z}}^v = \left( \mathcal{Z}^{v} + (\mathcal{Z}^{v})^T \right) / \sqrt{2}$
is a symmetric Gaussian random tensor field. (The $\sqrt{2}$ in the denominator accounts
for the variance reduction from averaging.)
Here $\mathcal{Z}^v$ is a white-noise random Gaussian tensor field; i.e.,
%
\begin{equation*}
\langle \mathcal{Z}^{v}_{\alpha\beta}(\mathbf{r},t)
\mathcal{Z}^{v}_{\gamma\delta}(\mathbf{r}',t') \rangle
= \delta_{\alpha\gamma} \delta_{\beta\delta} \,
\delta(\mathbf{r}-\mathbf{r}')\, \delta(t - t') \;\; .
\end{equation*}
%

\subsection{Stochastic Diffusion and Heat Fluxes}

The formulation of the multi-species stochastic diffusion and heat fluxes is complicated by the couplings among the species fluxes (cross-diffusion effects) and by the thermal diffusion contribution (Soret and Dufour effects).
The starting point for determining these fluxes
is the entropy production for a mixture, as formulated by de Groot and Mazur~\cite{DM_63} and by Kuiken~\cite{NewIrrevThermoBook},
which establishes the form of the thermodynamic forces and fluxes. We then use
the fluctuation-dissipation principle to formulate the corresponding noise terms.
Here we only need to consider the contributions of the heat flux and mass diffusion fluxes
to entropy production.
The entropy production also has a contribution due to the stress tensor, however,
due to the Curie symmetry principle \cite{DM_63}, fluxes and thermodynamic forces of different tensorial
character do not couple. As such, the stochastic flux in the momentum equation is
the same as for a single species fluid, as given by (\ref{StressTensorCorrelationEqn}).

The entropy production for a
multi-component mixture at rest, in the absence of external forces~\footnote{This contribution
is also zero if the external specific force
acting on each species is constant, as with a constant gravitational acceleration.} and chemistry,
is given by \cite{DM_63}:
\begin{eqnarray}
\EntropyProduction &=&
-\frac{1}{T^2} \HeatFlux' \cdot \nabla T
- \frac{1}{T} \sum_{i=1}^{N_s} \SpeciesFlux_i \cdot \nabla_T \mu_i \\
&=& -\frac{1}{T^2} \HeatFlux' \cdot \nabla T
- \frac{1}{T} \sum_{i=1}^{N_s-1} \SpeciesFlux_i \cdot \nabla_T \left(\mu_i - \mu_{N_s} \right ),
\label{eq:dGM1}
\end{eqnarray}
where $\mu_i$ is the chemical potential per unit mass of species $i$ and
\begin{equation}
\HeatFlux' = \HeatFlux - \sum_{k=1}^{N_s} h_k \SpeciesFlux_k
= \HeatFlux - \sum_{k=1}^{N_s-1} (h_k-h_{N_s}) \SpeciesFlux_k,
\end{equation}
where $h_k $ is the specific enthalpy of the $k^{th}$ component (see discussion in \ref{IdealGasMixturesSection}).
In other words, $\HeatFlux'$ is the part of the heat flux that is \emph{not} associated with mass diffusion.
Here, $\nabla_T$ is a gradient derivative taken holding temperature fixed, that is,
\[
\nabla_T  \; \mu_i(p,T,X_1, \ldots, X_{N_s-1}) = \nabla \mu_i -
\left(\frac{\partial \mu_i}{\partial T}\right)_{p,X_1, \ldots, X_{N_s-1}}\, \nabla T,
\]
where $X_k=n_k/\sum_{j=1}^{N_s} n_j$ are mole fractions, and $n_k$ are number densities.
The mole fraction for species $k$ is given in terms of the mass fractions by $X_k = ({\overline{m}}/{m_k}) Y_k$,
where $m_k$ is the
mass of a molecule of that species, and $\overline{m}=\left( \sum_{k=1}^{N_s}Y_k / m_k \right)^{-1}$ is the mixture-averaged molecular weight \cite{NewIrrevThermoBook}.  Note that only $N_s-1$ of the mass or mole fractions are independent.
%

The general form of the phenomenological laws
expresses the fluxes as linear combinations of thermodynamics forces, written in matrix form as
\[
\Jbar = \OnsagerMatrixB \Xbar
\qquad\mathrm{where}\qquad
\EntropyProduction = \Jbar^T \Xbar =  \Xbar^T \OnsagerMatrixB^T \Xbar.
\]
Here we use an overbar to denote the system expressed in terms of the first $N_s-1$ species.
From (\ref{eq:dGM1}) the fluxes $\Jbar$ and the thermodynamics forces $\Xbar$ are given by
\[
\Jbar =
\begin{bmatrix} \SpeciesFluxB \\ \HeatFlux'
\end{bmatrix}
\qquad \mathrm{and} \qquad
\Xbar = \begin{bmatrix}
- \frac{1}{T} \nabla_T ( \mu_i - \mu_{N_s})
\\
- \frac{1}{T^{2}}\nabla T
\end{bmatrix}
\]
respectively,  where $\SpeciesFluxB = [\SpeciesFlux_1,\ldots,\SpeciesFlux_{N_s-1}]^T$ is a vector of $N_s-1$ independent species mass fluxes.
By Onsager reciprocity the matrix of phenomenological coefficients is symmetric so we can write $\OnsagerMatrixB$ as
\[
\OnsagerMatrixB =
\begin{bmatrix}
{\Lbar} & \lbar
\\
{\lbar}^T & \ell
\end{bmatrix}  \;\;\;  ,
\]
where $\Lbar$ is a symmetric $N_s-1 \times N_s-1$ matrix that depends on the multicomponent flux diffusion coefficients, $\lbar$ is an $N_s-1$ component
vector that depends on the thermal diffusion coefficients, and the scalar $\ell$ depends on the
partial thermal conductivity (see \ref{IdealGasMixturesSection}).

Before discussing the form of the noise terms we will first recast $\OnsagerMatrixB$ in a slightly different form.
This form will facilitate comparison with the continuum transport literature (e.g.,~\cite{Giovangigli_99})  and lead
to a more efficient numerical algorithm.
We introduce
\[
\xibar = \Lbar^{-1} \lbar
\qquad \mathrm{and} \qquad
\zeta = \ell - \xibar^T \Lbar \xibar
\]
so that
\begin{equation}
{\OnsagerMatrixB} =
\begin{bmatrix}
{\Lbar} & {\Lbar \xibar}
\\
{\xibar^T{\Lbar}} & \zeta + \xibar^T {\Lbar} \xibar
\end{bmatrix}.
\label{eq:Lbar_xi}
\end{equation}
It is important to point out that this construction works even when $\Lbar$ is not invertible,
which happens when some of the species are not present. This is because $\xibar$ is always in the range of $\Lbar$.

We now want to establish the form of the stochastic fluxes in the fluctuating hydrodynamic equations.
Since the fluxes are white in space and time we can write them in the form
\[
\tilde{\Jbar}_\alpha = \Bcbar \Wbar^{(\alpha)}
\qquad \mathrm{where} \qquad
\widetilde{\Jbar}_\alpha = \begin{bmatrix}
\widetilde{\SpeciesFluxB}_\alpha
\\
\widetilde{\HeatFlux}'_\alpha
\end{bmatrix}
\qquad \mathrm{and} \qquad
\Wbar^{(\alpha)} = \begin{bmatrix}
\bar{\mathcal{Z}}^{(\SpeciesFlux;\alpha)}
\\
{\mathcal{Z}}^{(\HeatFlux';\alpha)}
\end{bmatrix}
\]
where $\alpha = x,y,z$ denotes spatial direction and $\bar{\mathcal{Z}}^{(\SpeciesFlux;\alpha)}
= [ \bar{\mathcal{Z}}^{(1;\alpha)},\ldots, \bar{\mathcal{Z}}^{(N_s-1;\alpha)} ]^T$
is a vector of independent Gaussian white noise terms,
that is,
\begin{eqnarray*}
\langle \bar{\mathcal{Z}}^{(i;\alpha)} (\mathbf{r},t)
\bar{\mathcal{Z}}^{(j;\beta)} (\mathbf{r}',t') \rangle
&=& \delta_{ij}\,
\delta_{\alpha\beta}\, \delta(\mathbf{r}-\mathbf{r}') \delta(t - t'), \\
\langle {\mathcal{Z}}^{(\HeatFlux';\alpha)} (\mathbf{r},t)
{\mathcal{Z}}^{(\HeatFlux';\beta)} (\mathbf{r}',t') \rangle
&=& \delta_{\alpha\beta}\, \delta(\mathbf{r}-\mathbf{r}') \delta(t - t'),
\end{eqnarray*}
and $\langle \bar{\mathcal{Z}}^{(\SpeciesFlux;\alpha)} {\mathcal{Z}}^{(\HeatFlux';\beta)} \rangle = 0$.

To satisfy fluctuation dissipation balance, we need \cite{Zarate_07,TernaryEquilibriumFluct}
\[
\Bcbar \Bcbar^T = 2 k_B \; \OnsagerMatrixB  \;\;\; .
\]
If we write the noise amplitude matrix
in the form
\[
\Bcbar =
\begin{bmatrix}
\Bbar & 0
\\
\xibar^T \Bbar & \sqrt{\zeta}
\label{eq:altnoiseform}
\end{bmatrix}
\]
then we obtain fluctuation-dissipation balance provided
\begin{equation}
 \Bbar \Bbar^T = 2 k_B \Lbar.
\label{BmatrixEqn}
\end{equation}
Note that the matrix $\Bbar$ is not uniquely-defined; for numerics we employ
the Cholesky factorization of $\Lbar$ to compute $\Bbar$, corresponding to choosing a
lower-triangular $\Bbar$.
From the above, the species diffusion flux noise is then,
\begin{equation}
\widetilde{\SpeciesFluxB}_\alpha  = \Bbar \; \bar{\mathcal{Z}}^{(\SpeciesFlux;\alpha)}
\label{MassDiffNoiseEqn}
\end{equation}
and the heat flux noise is,
\begin{eqnarray*}
\widetilde{\HeatFlux}_\alpha &=& \widetilde{\HeatFlux}_\alpha'
+ \bar{\mathbf{h}}^T \widetilde{\SpeciesFluxB}_\alpha \\
&=& \sqrt{\zeta} \mathcal{Z}^{(\HeatFlux';\alpha)} +
(\xi^T + \bar{\mathbf{h}}^T) \widetilde{\SpeciesFluxB}_\alpha,
\end{eqnarray*}
where $\bar{\mathbf{h}}$ is a vector with components $h_k - h_{N_s}$, the excess specific enthalpy.
The conservation of mass equation remains valid in the FNS equations so the sum of the species diffusion
noise terms for the full system must be zero.
Thus the stochastic mass flux for species $N_s$ is fixed by mass conservation.

For a given hydrodynamic system, the procedure for
computing the noise is
to determine $\Lbar$ in terms of mass diffusion coefficients from the phenomenological
law for $\SpeciesFlux_k$ and use (\ref{BmatrixEqn}) to compute $\Bbar$.
The phenomenological law for the heat flux can be used
to find expressions for $\xibar$ and $\zeta$. An example of this procedure is
given in Section~\ref{IdealGasMixturesSection} for a gas mixture.
More general non-ideal
fluid mixtures will be discussed in future work, including the relation of the above formulation
to the Stefan-Maxwell form of expressing the phenomenological relations between fluxes and thermodynamic forces \cite{NewIrrevThermoBook}.

\subsection{Full System Construction}

The form of the equations above requires that we distinguish a particular species, numbered $N_s$, which must be present
throughout the entire system. For many applications, this introduces an artificial requirement on the system
that is difficult to deal with numerically.
In this section we transform the reduced form with $N_s-1$ equations, used by de Groot and Mazur, to an equivalent full system construction.
It is noted in de Groot and Mazur that the Onsager reciprocal relations remain valid in the presence
of linear constraints such as (\ref{eq:det_sum_cond}).
In particular, we can consider the full system with $N_s+1$ equations (including thermal diffusion) with
the constraint $\sum_k (\mathcal{F}_k + \widetilde{\mathcal{F}}_k ) = 0$
by defining an augmented system that gives exactly the same entropy production.
In particular, we define an augmented Onsager matrix $\mathbf{L}$ of the form
\[
{\mathbf{L}} =
\begin{bmatrix}
\Lbar & -{\Lbar}\OnesVector \\
- \OnesVector^T {\Lbar} & \OnesVector^T\Lbar \OnesVector
\end{bmatrix}
\]
where $\OnesVector = [1,\ldots,1]^T$.  Here the final row gives $\SpeciesFlux_{N_s}$, the diffusion flux of the last species.
The extra row and column of ${\mathbf{L}}$ are fully specified by the
requirement that column sums vanish (a consequence of vanishing of the sum of
species fluxes) and the Onsager symmetry principle.

Using ${\mathbf{L}}$ we can write the phenomenological laws
for the full system as
\[
{\mathbf{J}} = {\OnsagerMatrix} {\mathbf{X}},
\]
where the fluxes ${\mathbf{J}}$ and thermodynamics forces ${\mathbf{X}}$ are given by
\[
{\mathbf{J}} =
\begin{bmatrix}
{\SpeciesFlux}
\\
\HeatFlux'
\end{bmatrix}
\qquad \mathrm{and} \qquad
{\mathbf{X}} = \begin{bmatrix}
- \frac{1}{T} \nabla_T  \mu
\\
-\frac{ \nabla T }{ T^2 }
\end{bmatrix}
\]
with
\[
{\OnsagerMatrix} =
\begin{bmatrix}
{\mathbf{L}} & {\mathbf{l}}
\\
{{\mathbf{l}}}^T & \ell
\end{bmatrix}
\qquad \mathrm{and} \qquad
{\mathbf{l}} =
\begin{bmatrix}
\lbar \\ - \OnesVector^T \lbar
\end{bmatrix},
\]
reflecting the fact that the Soret coefficients summed over all species vanishes.
Here, $\mu$ is a vector of all of the chemical potentials.
A direct computation shows that (\ref{eq:det_sum_cond}) gives
\[
\EntropyProduction
= \Jbar^T \Xbar
= {\mathbf{J}}^T {\mathbf{X}}
=  {\mathbf{X}}^T {\OnsagerMatrix} {\mathbf{X}}.
\]
Hence the full system form gives exactly the same entropy production as the original form.

Before constructing the noise for the full system,
we note that we can write the augmented Onsager matrix $\OnsagerMatrix$ in a form analogous to (\ref{eq:Lbar_xi}).
The key observation here is that
the construction of an extended ${\xi}$ remains valid because
${\mathbf{l}=\mathbf{L}\xi}$ is in the range of ${\mathbf{L}}$.
Note that ${\xi}$ is not uniquely determined.
We choose ${\xi}$ such that ${\xi}^T \OnesVector = 0$.
With these definitions, the Onsager matrix and associated noise term are given by
\begin{equation}
{\OnsagerMatrix} =
\begin{bmatrix}
{{\mathbf{L}}} & {{\mathbf{L}} {\xi}}
\\
{\xi}^T{{\mathbf{L}}} & \zeta + {\xi}^T {{\mathbf{L}}} {\xi}
\end{bmatrix}.
\label{eq:L_xi}
\end{equation}
Ottinger \cite{Ottinger_09} gives a derivation of this
form using the GENERIC formalism subject to the linear constraint $\sum_{k=1}^{N_s} Y_k=1$.
From (\ref{eq:L_xi}) we can then obtain the deterministic species flux
\begin{equation}
\SpeciesFlux = -\frac{1}{T} \mathbf{L} \left [ \nabla_T \mu + \frac{\xi}{T}\nabla T \right ]
\label{eq:F_Onsager}
\end{equation}
and the deterministic heat flux
\begin{equation}
\HeatFlux
= -\zeta \frac{\nabla T}{T^2}  +
(\xi^T + \mathbf{h}^T) \SpeciesFlux,
\label{eq:Q_Onsager}
\end{equation}
where $\mathbf{h}$ is the vector of specific enthalpies.

We can now construct the noise for the full system.  We note that since
$2 k_B \Lbar = \Bbar \Bbar^T$ we have that
\[
2 k_B {\mathbf{L}} = {\mathbf{B}} {\mathbf{B}}^T
\qquad \mathrm{where} \qquad
{\mathbf{B}} =
\begin{bmatrix}
\Bbar & 0 \\
- \OnesVector^T \Bbar & 0
\end{bmatrix}.
\]
In this form the species diffusion noise is given by
$\widetilde{\SpeciesFlux}_\alpha = {\mathbf{B}} \mathcal{Z}^{(\SpeciesFlux,\alpha)}$,
where $\mathcal{Z}^{(\SpeciesFlux,\alpha)}= [\bar{\mathcal{Z}}^{(\SpeciesFlux,\alpha)},0]$.
Although ${\mathbf{B}}$ is
of size $N_s \times N_s $, only $N_s-1$ noise terms are needed because the last column of
${\mathbf{B}}$ is identically zero.
Note also that the last row is chosen so that the sum of the noise terms over all species vanishes.
We can now define the noise matrix for species diffusion, ${\mathcal{B}}$,
such that fluctuation-dissipation balance is obeyed, ${\mathcal{B}}{\mathcal{B}}^T = 2 k_B \OnsagerMatrix$, namely,
\begin{equation}
{\mathcal{B}} =
\begin{bmatrix}
{\mathbf{B}} & 0
\\
{\xi}^T {\mathbf{B}} & \sqrt{\zeta}
\end{bmatrix}.
\label{eq:finalnoiseform}
\end{equation}
The augmented stochastic heat flux is thus given by
\begin{equation*}
\widetilde{\HeatFlux}_\alpha
= \sqrt{\zeta} \mathcal{Z}^{(\HeatFlux';\alpha)} +
(\xi^T + \mathbf{h}^T) \widetilde{\SpeciesFlux}_\alpha,
\end{equation*}
in analogy (and fluctuation-dissipation balance) with the deterministic heat flux (\ref{eq:Q_Onsager}).
This form is identical to that given by Ottinger~\cite{Ottinger_09}
and we use it in the next section to establish the relationship
between the Onsager matrix and deterministic transport models.
Note that ${\zeta} \geq 0$ since $\OnsagerMatrixB $
must be positive definite  while ${\OnsagerMatrix}$ must be positive semi-definite.

Finally, the methodology can also be applied when not all species are present.
Rows and columns of ${\mathbf{L}}$ corresponding to missing species are identically zero.
By applying a suitable permutation matrix $P$ to obtain $\check{\mathbf{L}} = P {\mathbf{L}} P^T$
we can arrange for the missing species to be the last rows and columns of $\check{\mathbf{L}}$.
If $m$ species are present then the upper $m \times m$ block has rank $m-1$ with the structure discussed above.
For $\check{{\xi}} = P {{\xi}}$ the first $m$ elements sum to zero and the
last $N_s - m$ elements are equal to zero.
We note that although the extension of the formalism is straightforward,
some care is needed to prevent numerical roundoff error from spuriously generating small amounts of absent species.

\subsection{Gas Mixtures}\label{IdealGasMixturesSection}

The hydrodynamic properties of a fluid are fixed by its thermodynamic functions (e.g., equation of state) and its
transport properties.
This section summarizes these relations for a multi-species mixture of gases, following the notation
in \cite{Giovangigli_99}. The ideal gas equation of state is
\begin{equation}
p = R_u T \sum_{k=1}^{N_s} \frac{\rho_k}{W_k} =
\rho R_u T \sum_{k=1}^{N_s} \frac{Y_k}{W_k} = \frac{\rho R_u T}{\overline{W}},
\end{equation}
where $R_u = k_B N_A$ is the universal gas constant, $N_A$ is Avogadro's number,
the molecular weight of the $k$-th species is $W_k = m_k N_A$, and $\overline{W} = \overline{m} N_A$ is the mixture-averaged molecular weight.

The total specific energy is
\begin{equation}
E = \frac{1}{2}|{\bf v}|^2 +  e,
\end{equation}
where $e$ is the specific internal energy. For an ideal gas mixture we can write,
\begin{equation}
e\left( T, Y_k \right) = \sum_{k=1}^{N_s} Y_k e_k(T),
\end{equation}
where $e_k$ is the specific internal energy of the $k$-th species.
Similarly, we can write the specific and partial enthalpies as
\begin{equation}
h = e + \frac{p}{\rho} = \sum_{k=1}^{N_s} Y_k h_k(T)
\qquad\mathrm{and}\qquad
h_k = e_k + \frac{R_u}{W_k} T.
\label{EnthalpyEquation}
\end{equation}
The specific heats at constant volume and pressure for the mixture are:
%
\begin{eqnarray*}
c_v(T) = \left( \frac{\partial e}{\partial T} \right)_{Y_k, v} = \sum_{k=1}^{N_s} Y_k c_{v,k}(T);\\
c_p(T) = \left( \frac{\partial h}{\partial T} \right)_{Y_k, p} = \sum_{k=1}^{N_s} Y_k c_{p,k}(T).
\end{eqnarray*}
Given $c_{v,k}$ and $c_{p,k}$ one obtains $e_k(T)$ and $h_k(T)$ by integration.
For a calorically perfect gas, $c_{v,k}$ and $c_{p,k}$ are constants, and for a thermally perfect gas they are usually expressed as polynomial expressions in $T$.

For an ideal gas the chemical potential per unit mass can be written as,
\[
\mu_i = \frac{R_u T}{W_i} (\ln X_i + \ln p) + f(T),
\]
where $f(T)$ is a function only of temperature. Recalling that $\mu$ represents
the vector of $\mu_i$ then we have
\begin{eqnarray*}
\nabla_T \; \mu &=&  {R_u T}{\cal{W}}^{-1}  {\cal{X}}^{-1} \nabla X +
\frac{R_u T}{p} \mathcal{W}^{-1} \OnesVector \nabla p \\
&=& \frac{R_u T}{\overline{W}} {\cal{Y}}^{-1} \nabla X + \frac{\overline{W}}{\rho} \mathcal{W}^{-1} \OnesVector \nabla p,
\end{eqnarray*}
where $X$ and $Y$ are vectors of mole fractions and mass fractions, respectively,
$\cal{X}$, $\cal{Y}$ and $\cal{W}$
are diagonal matrices of mole fractions, mass fractions and molecular weights, and $\mathbf{u}$ is vector of all ones.

We will use this form
to relate the transport coefficients to the noise amplitude matrix.
Software libraries, such as EGLIB~\cite{EGLIB},
used to compute these transport coefficients
typically express fluxes in terms of gradients of $X$, $p$ and $T$
rather chemical potential.
In particular, these packages typically compute: a matrix of
multicomponent flux diffusion coefficients, $\mathbf{C}$,
a vector of thermal diffusion coefficients $\theta$ or rescaled thermal diffusion ratios,
$\tilde{\chi}$; and either thermal conductivity $\lambda$ or
partial thermal conductivity $\hat{\lambda} = \lambda + \frac{p}{T} \tilde{\chi}^T \mathcal{X} \theta$.
Here, $\theta$ and $\tilde{\chi}$ are related by
\[
\mathbf{C} \mathcal{X} \tilde{\chi} = \rho \mathcal{Y} \theta
\]
Computation of $\tilde{\chi}$ and $\lambda$ is more computationally efficient than
computation of $\theta$ and $\hat{\lambda}$ so we will focus on relating ${\OnsagerMatrix}$
as given in (\ref{eq:L_xi}) to the diffusion fluxes expressed in terms of $\mathbf{C}$,
$\tilde{\chi}$ and $\lambda$.

In terms of these variables we then have
\begin{eqnarray}
\SpeciesFlux &=&
- \mathbf{C} \left( d + \mathcal{X} \tilde{\chi} \frac{\nabla T}{T} \right);
\label{GasSpeciesFluxEqn}
\\
\HeatFlux' = \HeatFlux - h^T \SpeciesFlux  &=&
 -\lambda \nabla T + R_u T \tilde{\chi}^T \mathcal{W}^{-1} \SpeciesFlux
 \label{GasHeatFluxEqn}
\end{eqnarray}
For an ideal gas, the diffusion driving force \cite{NewIrrevThermoBook} is
\[
{d} = {\nabla} X + \left( X - Y \right) \frac{{ \nabla} p}{p}.
\]
In (\ref{GasSpeciesFluxEqn}) and (\ref{GasHeatFluxEqn}) we can replace
$d$ with $\hat{ d}$ where
\[
\hat{ d} = { \nabla} X + X \frac{{\bf \nabla} p}{p}.
\]
These forms are equivalent because $\mathbf{C} Y = 0$ and $\theta^T Y = 0$.
The additional term in ${d}$ is to enforce ${d}^T \OnesVector = 0$ by adding to
$\hat{{d}}$ an appropriate element in the null space of $\mathbf{C}$.

By comparison, in the phenomenological laws,
${\mathbf{J}} = {\OnsagerMatrix} {\mathbf{X}}$,
the flux is given by
\begin{eqnarray*}
\SpeciesFlux &=& - {\mathbf{L}} \left( \frac{1}{T} \nabla_T \mu
+ {\xi} \frac{1}{T^2} \nabla T \right) \\
&=& - {\mathbf{L}} \left( \frac{R_u}{\overline{W}} \mathcal{Y}^{-1} \nabla X
+ R_u \mathcal{W}^{-1} \frac{\nabla p}{p}
+ \frac{1}{T^2} {\xi} \nabla T \right) \;\;\; .
\end{eqnarray*}
By matching the $\nabla X$ terms
we have,
\begin{equation}
{\mathbf{L}} = \frac{\overline{W}}{R_u} \mathbf{C} \mathcal{Y}.
\label{Giovangigli_L}
\end{equation}
A bit of algebra gives the same result for the $\nabla p$ term, which is
the baro-diffusion contribution. Note that, in general, the $\nabla X$ and $\nabla p$ terms
will yield the same result since the baro-diffusion contribution is of thermodynamic origin
and thus it does not have an associated transport coefficient~\cite{Landau_59}.
From the $\nabla T$ term,
\[
{\xi} = \frac{R_u T}{\bar{W} } \mathcal{Y}^{-1} \mathcal{X} \tilde{\chi} =
R_u T \mathcal{W}^{-1}  \tilde{\chi},
\]
which corresponds to the Soret term in the species diffusion equations.

Similarly, in the phenomenological laws,
using the expression for heat flux we have,
\[
\HeatFlux' = \HeatFlux - h^T \SpeciesFlux = -\frac{\zeta}{T^2} \nabla T + \xi^T \SpeciesFlux ,
\]
which by comparison to (\ref{GasHeatFluxEqn}) gives the relation
\begin{equation}
\zeta = T^2 {\lambda}.
\end{equation}

\section{Numerical Scheme}
\label{sec:numerical}

The numerical integration of (\ref{eqn:spec})-(\ref{eqn:energy}), (\ref{eqn:cont})
is based on a method of lines approach in which
we discretize the equations in space and then use an ODE integration algorithm
to advance the solution.
Here we use
the low-storage third-order Runge-Kutta (RK3) scheme previously
used to solve the single and two-component FNS equations \cite{Donev_10}, using the
weighting of the stochastic forcing proposed by Delong {\it et al.} \cite{TemporalIntegratorsPaper2013}.
We can write the governing equations in the following form:
\begin{equation}
\frac{\partial \U}{\partial t} = -
\nabla \cdot \FH - \nabla \cdot \FD - \nabla \cdot \FS + \mathbf{H} \equiv \mathbf{R}(\U, Z)
\end{equation}
where $\U = \left[\rho, \rho Y_k, \rho \mathbf{v}, \rho E\right]^{T}$
is the set of conservative variables, $\FH$, $\FD$, and $\FS$ are the
hyperbolic, diffusive and stochastic flux terms, respectively and $\HH$ is
a forcing term.
Here, $\mathbf{R}$ is a shorthand for the right hand side of the equation used later for describing
the temporal discretization scheme and $Z$ is a spatio-temporal discretization of
the random Gaussian fields $\mathcal{Z}$ used to construct the noise.

The fluxes are given by
\begin{equation}
\FH =
\begin{bmatrix}
\rho \mathbf{v} \\
\rho \mathbf{v} Y_k  \\
\rho \mathbf{vv}^T + p\mathbf{I} \\
\rho \mathbf{v} (E + p)
\end{bmatrix}
\qquad ; \qquad
\FD =
\begin{bmatrix}
0 \\
{\SpeciesFlux} \\
{\StressTensor}  \\
{\HeatFlux} + {\StressTensor} \cdot \mathbf{v}
\end{bmatrix}
\qquad ; \qquad
\FS =
\begin{bmatrix}
0 \\
\widetilde{\SpeciesFlux} \\
\widetilde{\StressTensor}  \\
\widetilde{\HeatFlux} + { \widetilde{\StressTensor}} \cdot {\bf v}  \\
\end{bmatrix}.
\end{equation}
Here we consider only a gravitational
source term
$\HH = [0, 0, \rho {\bf g} , \rho {\bf g \cdot \mathbf{v}}]$;
however, in a more general case
$\HH$ can include both deterministic and stochastic forcing terms (e.g., chemical reactions).
Thus, for $N_s$ species, $k = 1, \ldots, N_s$ there is a total of $5 + N_s$ governing equations in three dimensions.
Note that for a single-species fluid the equation for $\rho_1$ and $\rho$ are identical.
\footnote{We note that technically, we to not need to solve a separate continuity equation since
$\rho =\sum_k \rho Y_k$. We do so here simply for diagnostic purposes.}

\subsection{Spatial discretization}

The spatial discretization uses a finite volume representation with
cell spacings in the $x$-, $y$- and $z$-directions given by
$\Delta x$, $\Delta y$ and $\Delta z$.
We let $\U_{ijk}^n$ denote the average value of $\U$ in cell-$ijk$ at time step $n$.
To ensure that the algorithm satisfies discrete fluctuation-dissipation
balance, the spatial discretizations are done using centered discretizations (see Donev {\it et al.}
\cite{Donev_10}).

To obtain the hyperbolic fluxes we first compute
the primitive variables, $\rho$, $Y_k$, $\mathbf{v}$, $T$, and $p$
from the conserved variables at cell centers.
These values are then interpolated to cell faces
using a PPM-type \cite{Colella_84} spatial interpolation.
For example, for temperature we set
\begin{equation}
T_{i+1/2,j,k}^n = \frac{7}{12} \left(T_{i+1,j,k}^n + T_{i,j,k}^n  \right ) - \frac{1}{12} \left(T_{i-1,j,k}^n + T_{i+2,j,k}^n  \right )
\;\; .
\end{equation}
From these interpolants we evaluate the flux terms at the face.
The divergence of the fluxes is then computed as $\Detoc \; \FH$
where $\Detoc$ is the standard
discrete divergence operator that computes the cell-centered divergence
of a field defined on cell faces.

The computation of the diffusive and stochastic terms is a bit more complex.
The evaluation of the deterministic
heat flux and the species diffusion terms is done in a straightforward
fashion using the face-based operators and simple arithmetic averages to compute
transport coefficients at cell faces.
However, a complication arises because the viscous stress tensor $\StressTensor$ uses a symmetrized gradient, namely
\[
\StressTensor = -\eta (\nabla \mathbf{v} + (\nabla \mathbf{v})^T) - (\kappa - \frac{2}{3} \eta )
( \mathbf{I} \; \nabla \cdot \mathbf{v} ) \;\; .
\]
Standard discretizations of the stress tensor in this form do not satisfy
a discrete fluctuation dissipation balance.  More precisely, they lead to a weak correlation
between velocity components at equilibrium.
These problems stem from the fact that the concept of a symmetric stress tensor
does not have a natural expression on a cell-centered grid as employed here \cite{Donev_10}.
By contrast, if a staggered grid is used to handle the momentum equation, it is straightforward to
construct a symmetric stochastic stress tensor using straightforward centered second-order staggered difference operators \cite{Balboa2012}. The staggered grid discretization is particularly useful for incompressible flow; here we consider
the full compressible equations and focus on cell-centered grids.

Extending the development in \cite{Donev_10} to the case of variable viscosity,
we first rewrite the viscous term in the form,
\begin{equation}
\nabla \cdot
\StressTensor = - \nabla \cdot \left( \eta \nabla \mathbf{v} \right) -
\nabla \left[ (\kappa + \frac{1}{3} \eta ) \; (\nabla \cdot \mathbf{v}) \right]
+ \left[ (\nabla \eta )  ( \nabla \cdot \mathbf{v})
- (\nabla \eta) \cdot (\nabla \mathbf{v})^T \right].
\label{eq:altstress}
\end{equation}
Observe that the last two terms only involve first derivatives of $\mathbf{v}$ and, in fact,
vanish completely when $\eta$ is a constant and were not omitted in \cite{Donev_10}.
We note that this rewriting of the stress tensor can be written in a conservative form in which
there are cancellations in the last two terms,
\[
\StressTensor = - \nabla \cdot \left( \eta \nabla \mathbf{v} \right)-
\nabla \cdot \left [ (\kappa + \frac{1}{3} \eta ) \; \mathbf{I} \; (\nabla \cdot \mathbf{v}) \right]
+ \left[ \nabla \cdot (\eta \;  \mathbf{I} \; (\nabla \cdot \mathbf{v}))
- \nabla \cdot (\eta (\nabla \mathbf{v})^T) \right].
\]
Our spatial discretization follows this conservative form and thus ensures discrete conservation of momentum.

We use different discretizations of the different terms in equation (\ref{eq:altstress}).
For the first term, we approximate
\begin{equation}
\nabla \cdot \left( \eta \nabla \mathbf{v} \right)
\approx
\Gctoe \left( \eta  \; \Detoc \mathbf{v} \right),
\label{eq:ten1}
\end{equation}
where $\Gctoe$ defines normal gradients at cell faces from cell centered values.
Here, we average adjacent cell-centered values of $\eta$ to edges.
For the remaining terms we use a nodal (corner) based discretization. For example we approximate
\begin{equation}
\nabla \left[ (\kappa + \frac{1}{3} \eta ) \; (\nabla \cdot \mathbf{v}) \right]
\approx
\Gcton \left[ (\kappa + \frac{1}{3} \eta ) \; \Dntoc \mathbf{v} \right],
\label{eq:ten2}
\end{equation}
where $\Dntoc$ uses nodal values of a field to compute the divergence at
cell centers and $\Gcton$ computes gradients at corner nodes from cell-centered values.
Again, the discretizations are standard second-order difference approximations.
Here, coefficients are computed by averaging cell-centered values at all eight
adjacent cells centers to the node.
We also discretize the last terms in (\ref{eq:altstress}) using nodal discretizations based on the
conservative form,
\begin{equation}
\left[ \nabla \cdot (\eta \;  \mathbf{I} \; (\nabla \cdot \mathbf{v}))
- \nabla \cdot (\eta (\nabla \mathbf{v})^T) \right]
\approx
\Gcton \left(  \eta \; \Dntoc \mathbf{v} \right)
+ \Dntoc \left( \eta \; (\Gcton \mathbf{v})^T \right),
\label{eq:ten3}
\end{equation}
noting that the second-order derivative terms cancel at the discrete level just as they do
in the continuum formulation, leaving only first-order differences when the two terms are combined.

With these definitions, we have that $\Detoc = -\GctoeT$ and $\Dntoc = -\GctonT$, i.e.,
both
the nodal and face-based discrete divergence and gradient operators are discretely
skew-adjoint.  These skew-adjoint properties are important for numerically satisfying discrete
fluctuation-dissipation balance.
The viscous heating contribution to the energy equation,
$\nabla \cdot (\StressTensor \cdot \mathbf{v})$ is evaluated using face centered values of $\StressTensor$ multiplied
by an arithmetic averge of $\mathbf{v}$ to faces from cell centers.  The terms of $\StressTensor$ corresponding to (\ref{eq:ten1})
are defined on faces; the terms corresponding to (\ref{eq:ten2}) and (\ref{eq:ten3}) are computed by averages of corner values
to faces and forming
$\StressTensor \cdot \mathbf{v}$ at faces then computing the divergence of the fluxes using $\Detoc$.

The noise terms in the momentum equation that represent the stochastic
stress tensor need to respect the correlation structure given in
(\ref{StressTensorCorrelationEqn}).
In addition, the discrete treatment of the noise needs to match the discretization
of the deterministic stress tensor. In particular, they need to use the same
discrete divergence.  This, combined with the skew adjoint construction of the
gradient operators, is needed for fluctuation-dissipation balance.
For that reason, we generate noise terms for the first two terms in
(\ref{eq:altstress}) separately.  No stochastic terms are added for the last two
terms because they only involve first derivatives of $\mathbf{v}$.

The stochastic stress tensor is expressed as $\widetilde{\StressTensor}
= \widetilde{\StressTensor}^{(f)} + \widetilde{\StressTensor}^{(n)}$.
The term $\widetilde{\StressTensor}^{(f)}$ corresponds
to the $\nabla \cdot \left( \eta \nabla \mathbf{v} \right)$ contribution to the dissipative (viscous) flux; at a face we form it as
\[
\widetilde{\StressTensor}_{i+\half,j,k}^{(f)} =
\sqrt{
2 k_B (\eta T)_{i+\half,j,k} } \mathfrak{S} Z^{(v,x)},
\]
where
\begin{equation}
(\eta T)_{i+\half,j,k}=(\eta_{i,j,k}T_{i,j,k}+\eta_{i+1,j,k}T_{i+1,j,k})/2,
\label{eq:c2f_averaging}
\end{equation}
and $Z^{(v,x)}$ are three-component,
independent face-centered standard Gaussian random variables and
\begin{equation}
\mathfrak{S} = \frac{1}{\sqrt{\Delta x ~ \Delta y ~  \Delta z ~ \Delta t}}
\label{eq:scale}
\end{equation}
is a scaling due to the $\delta $ function correlation in space and time of the noise,
see \cite{Donev_10,TemporalIntegratorsPaper2013} for a more precise derivation.
Other faces are treated analogously and the resulting stochastic momentum fluxes are differenced
using the discrete divergence $\Detoc$.

The stochastic flux corresponding to the contribution
$ \nabla \left[ (\kappa + \frac{1}{3} \eta ) \; (\nabla \cdot \mathbf{v}) \right]$
in the dissipative flux is generated at corner nodes \cite{Donev_10}. Namely,
\[
\widetilde{\StressTensor}_{i+\half,j+\half,k+\half}^{(n)} =
\sqrt{
2 k_B \left [(\kappa + \frac{1}{3}
 \eta)T \right ]_{i+\half,j+\half,k+\half} }~\mathfrak{S} Z^{(v,n)},
\]
where $Z^{(v,n)}$ are three-component,
independent node-centered standard Gaussian random variables.
Note that the coefficients at the corner nodes are averages over the eight cells adjacent to the
node, analogously to (\ref{eq:c2f_averaging}).
The divergence of these nodal fluxes is computed using the discrete divergence operator $\Dntoc$.
The viscous heating contribution from the stochastic stress is computed analogously
to the deterministic contribution described above.

The noise terms for the species and energy equation are generated in the full-system form
${\mathcal{B}}$ using the expressions written in terms of ${\mathbf{L}}$, ${\xi}$,
and $\zeta$.
Here we use the particular form of these expressions given for gas mixtures.
In particular, for edge $i+\half,j,k$ we define
\[
\mathbf{L}_{i+\half,j,k} =
\left( \frac{\overline{W}_{i,j,k}+\overline{W}_{i+1,j,k}}{2 R_u} \right)
\left( \frac{ (C \mathcal{Y})_{i,j,k}+ (C \mathcal{Y})_{i+1,j,k}}{2} \right)
\]
and obtain ${\mathbf{B}}$ by forming the Cholesky decomposition of
$ \mathbf{L}_{i+\half,j,k} $,
\[
\mathbf{B}_{i+\half,j,k} \mathbf{B}_{i+\half,j,k}^T =
2 k_B \mathbf{L}_{i+\half,j,k}  \;\;\;  .
\]
The stochastic flux for species is then given by
\[
\widetilde{\mathcal{F}}_{i+\half,j,k} = \mathbf{B}_{i+\half,j,k} \mathfrak{S} Z_{i+\half,j,k}^{(F,x)},
\]
where $Z^{(F,x)}$ are face-centered independent standard Gaussian random variables.
Stochastic fluxes on other edges are constructed analogously and the divergence is
computed with $\Detoc$.

We then define
\[
{\xi}_{i+1/2,j,k} =
\frac{R_u (T_{i,j,k}+T_{i+1,j,k})}{2} \mathcal{W}^{-1}
\left( \frac { \tilde{\chi}_{i,j,k}+\tilde{\chi}_{i+1,j,k}}{2} \right).
\]
The noise term, $\widetilde{Q}_x$, in the energy flux is then
\[
\widetilde{Q}_{i+\half,j,k} = \sqrt{k_B (\zeta_{i,j,k}+\zeta_{i+1,j,k})} \mathfrak{S} Z^{(Q,x)}
+ \left( {\xi}_{i+\half,j,k}^T + h_{i+\half,j,k}^T \right) \widetilde{\mathcal{F}}_{i+\half,j,k},
\]
where $Z^{(Q,x)}$ are face-centerd independent standard Gaussian random variables.
Here, $h_{i+\half,j,k}$ is obtained by evaluating the specific enthalpies at the temperature
\[
  T_{i+\half,j,k}= (T_{i,j,k}+T_{i+1,j,k}) / 2,
\]
and the same face-centered value of $\left( {\xi}_{i+\half,j,k} + h_{i+\half,j,k} \right)$ is used to weight the
contribution of mass fluxes to the heat flux for both the deterministic and the stochastic fluxes.

\subsection{Temporal discretization}

The temporal discretization uses the low-storage third-order Runge-Kutta (RK3) scheme previously
discussed in Donev {\it et al.} \cite{Donev_10} using the weights specified in
\cite{TemporalIntegratorsPaper2013}. With this choice of weights, the temporal
integration is weakly second-order accurate for additive noise (e.g., the linearized
equations of fluctuating hydrodynamics \cite{Zarate_07}).

The RK3 scheme involves three stages, which can be summarized as follows:
\begin{eqnarray}
\U_{i,j,k}^{n+1/3} &=& \U_{i,j,k}^n + \Delta t \mathbf{R}(\U^n,Z_1); \nonumber \\
\U_{i,j,k}^{n+2/3} &=& \frac{3}{4}\U_{i,j,k}^{n} +
\frac{1}{4} \left [ \U_{i,j,k}^{n+1/3} + \Delta t \mathbf{R}(\U^{n+\frac{1}{3}},Z_2) \right ]; \\
\U_{i,j,k}^{n+1} &=& \frac{1}{3}\U_{i,j,k}^{n} + \frac{2}{3} \left [ \U_{i,j,k}^{n+2/3} +
\Delta t \mathbf{R}(\U^{n+\frac{2}{3}},Z_3) \right ], \nonumber
\end{eqnarray}
where the $Z_i$ denote the random fields used in each stage of the integration.
To compute the weights for each stage, we generate
two sets of normally distributed independent Gaussian fields, $Z^A$ and $Z^B$, and
set
\begin{eqnarray*}
Z_1 &=&  Z^A + \beta_1 Z^B; \\
Z_2 &=&  Z^A +  \beta_2  Z^B; \\
Z_3 &=&  Z^A+  \beta_3  Z^B,
\end{eqnarray*}
where $\beta_1 = (2 \sqrt{2}+ \sqrt{3})/5 $, $\beta_2 = (-4 \sqrt{2}+ 3 \sqrt{3})/5$, and
$\beta_3 = (\sqrt{2} - 2 \sqrt{3})/10$.

\subsection{Boundary conditions}

In addition to periodicity, our implementation of the methodology described above
supports three boundary conditions.
The first is a specular wall at which the normal velocity vanishes and
the other velocity components,
mole fractions and temperature satisfy homogeneous Neumann boundary conditions.
A second type
of boundary condition is a no slip, reservoir wall at which the normal velocity vanishes
and the other velocity components, mole fractions and temperature satisfy inhomogeneous
Dirichlet boundary conditions.
The third boundary condition is a variant of the no slip condition
for which the wall is impermeable to species so that
the normal derivative of mole fraction vanishes.
When a Dirichlet condition is specified for a given quantity, the corresponding diffusive
flux is computed as a difference of the cell-center value and the value on the boundary.
In such cases the corresponding stochastic flux is multiplied by $\sqrt{2}$ to ensure
discrete fluctuation-dissipation balance, as explained in detail \cite{Balboa2012,DonevLowMach2013}.

\section{Numerical Results}
\label{sec:results}

In this section we describe several test problems
that demonstrate the capabilities of the numerical methodology.
The first two examples serve as validation that the methodology produces the
correct fluctuation spectra in both equilibrium and non-equilibrium settings.  The other two
examples illustrate the type of phenomena that can occur in multicomponent
systems.

\subsection{Equilibrium mixture of gases}

We start with equilibrium simulations of non-reacting, multi-species mixtures,
specifically, four noble gases (see Table \ref{tab:example1}).
The hard sphere model was used with the ideal gas equation of state
and $c_{v,k} = 3 k_B/2 m_k$.
For the hard sphere transport coefficients, $\eta$ and ${\lambda}$ were evaluated
using the dilute gas formulation in~\cite{HCB_54}; for $\widetilde{\chi}$ it was
more convenient to use the formulation in~\cite{Valk_63}.
Finally, the binary diffusion coefficients, as formulated in~\cite{HCB_54},
were used to obtain $\mathbf{C}$ using a numerically efficient iterative method
from~\cite{Giovangigli_99}.

The system was initialized at rest with
pressure $p = 1.01\times 10^6~\mathrm{dyn/cm}^2$ and temperature $T = 300$~K.
The density was $\rho = 4.83\times 10^{-4}~\mathrm{g/cm}^3$ with
initial mass fractions of $Y_k=0.25$ for each species, leading to a wide range in
mole fractions, as shown in the table.  The simulations were run in a $64^3$ domain
with periodic boundary conditions,
cell dimensions of $\Delta x = \Delta y = \Delta z = 8 \times 10^{-6}$~cm,
and a time step of
$\Delta t = 10^{-12}$~s, corresponding to an acoustic Courant number \cite{Donev_10} of between 0.15 and 0.2.
At these conditions, the fluctuations are fairly significant with instantaneous
variations in $\rho$ within the domain of the order of 10\%.

\begin{table}
\begin{tabular}{|c|l|c|c|c|c|}
\hline
$~k~$ & Species  & Molecular Weight & Diameter (cm) & $Y_k$ & $X_k$ \\
\hline
1 & Helium & 4.0026 & 2.18 $\times 10^{-8}$ & 0.25 & 0.7428 \\
\hline
2 & Neon & 20.1797 & 2.58 $\times 10^{-8}$ & 0.25 & 0.1473 \\
\hline
3 & Argon &  39.9480 & 3.63 $\times 10^{-8}$ & 0.25 & 0.0744 \\
\hline
4 & Krypton &  83.8000 & 4.16 $\times 10^{-8}$ & 0.25 & 0.0355 \\
\hline
\end{tabular}
\caption{Molecular properties for the equilibrium test case. }
\label{tab:example1}
\end{table}

Simulations were initially run for an equilibration time of 40000 time steps
and then the run continued for approximately
500000 additional time steps, with data collected every 10 time steps.
The data from the spatial computational grid
was then Fourier transformed in 3D and pair-wise correlations were computed for
each wave number and averaged in time.
These static structure factors were normalized by the equilibrium values (see Appendix \ref{AppendixEq})
except for the case of correlations that are zero at equilibrium.
For those correlations the normalization
used the corresponding variances, for example,
$|\langle (\delta \hat{\rho}) (\delta \hat{T}^*) \rangle |$
is normalized by
$\sqrt{ \langle (\delta\hat{\rho}) (\delta\hat{\rho}^*) \rangle
\langle (\delta\hat{T}) (\delta\hat{T}^*) \rangle}$ \cite{Donev_10}.

\begin{figure}
(a)
\includegraphics[width=2in]{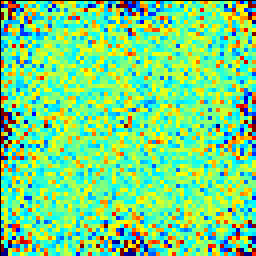}
(b)
\includegraphics[width=2in]{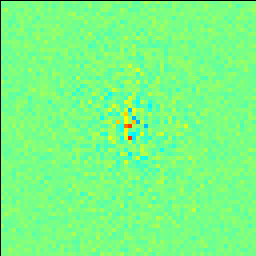}
\hspace{.75in}

(c)
\includegraphics[width=2in]{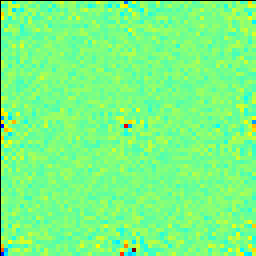}
(d)
\includegraphics[width=2in]{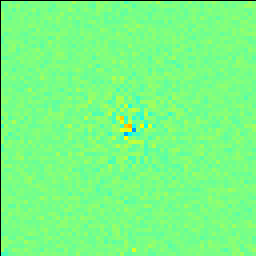}
\hspace{.75in}

(e)
\includegraphics[width=2in]{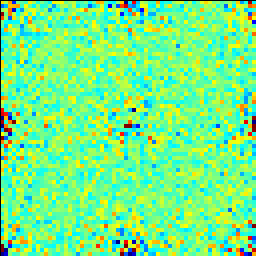}
(f)
\includegraphics[width=2in]{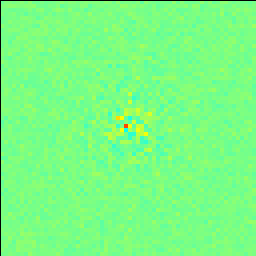}
\includegraphics[width=.75in]{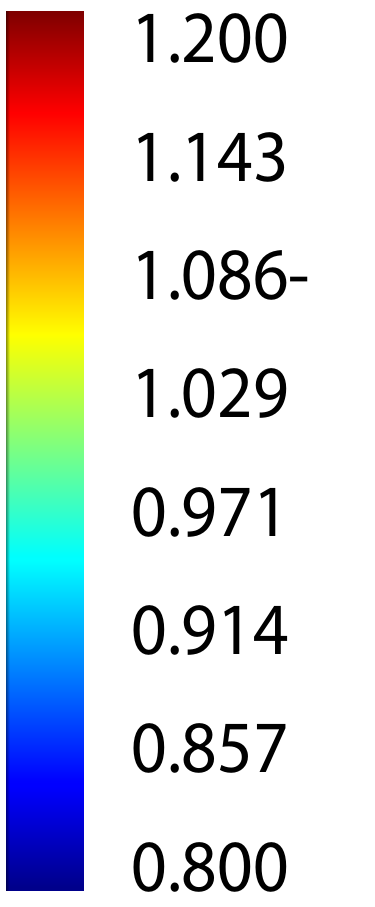}
\caption{(Color online) Static structure factors.
(a) $\langle (\delta\hat{\rho}) (\delta\hat{\rho}^*)\rangle)$,
(b) $\langle (\delta\widehat{J_x}) (\delta\widehat{J_x}^*)\rangle$,
(c) $\langle (\delta\widehat{\rho E}) (\delta\widehat{ \rho E}^*)\rangle$,
(d) $\langle (\delta\hat{\rho_1}) (\delta\hat{\rho_1}^*)\rangle$,
(e) $\langle (\delta\hat{\rho_4}) (\delta\hat{\rho_4}^*)\rangle$,
(f) $\langle (\delta\hat{T}) (\delta\hat{T}^*)\rangle$.
Data range is set to $\pm 20\%$ of the theoretical value of unity, as shown in the color bar.
The wave number domain represented here is
$[-3.93 \times 10^7,3.93 \times 10^7] \times [-3.93 \times 10^7,3.93 \times 10^7] $ in units
of cm$^{-1}$.}
\label{fig:variances}
\end{figure}

In Figure \ref{fig:variances} we present selected static structure factors from the simulation.  In
each case, the color scale is adjusted to represent $\pm 20$\% of the equilibrium value, which is
independent of wave number.
The simulations show excellent agreement with the theoretical values.
The structure factor for $\rho$, given by
$\langle (\delta\hat{\rho}) (\delta\hat{\rho}^*) \rangle$, is the noisiest.
This occurs because the continuity equation, (\ref{eqn:cont}),
does not contain either a diffusive term or a stochastic flux so density fluctuations
are solely driven by velocity fluctuations in the hyperbolic flux.
Nevertheless, the (unnormalized) variance of density (i.e., the average static structure factor over all wavenumbers) is
$5.7904\times 10^{-11}~\mathrm{g}^2/\mathrm{cm}^6$,
which is within 0.07\% of the analytic value of $5.7942 \times 10^{-11}~\mathrm{g}^2/\mathrm{cm}^6$.
The other relatively noisy structure factor is $\langle (\delta\hat{\rho}_4) (\delta\hat{\rho}_4^*) \rangle$,
which is a result of the relatively low mole fraction which makes $\rho_4$ noisy with
over 40\% instantaneous variation. Here again, the average variance is correct
to within 0.15\%.
Note that the grid-based finite-volume methods we employ here are neither
translationally (Gallilean) invariant nor rotationally invariant, and this leads
to non-isotropic structure factors for finite time step sizes (i.e., to
non-isotropic spatial truncation errors), particularly for high wave numbers.

\begin{figure}
(a)
\includegraphics[width=2in]{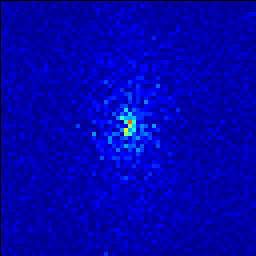}
(b)
\includegraphics[width=2in]{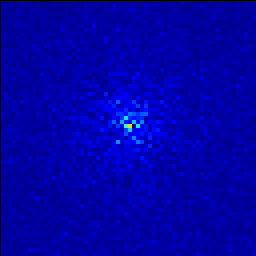}
\hspace{.75in}

(c)
\includegraphics[width=2in]{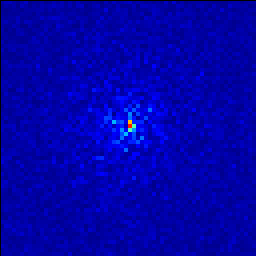}
(d)
\includegraphics[width=2in]{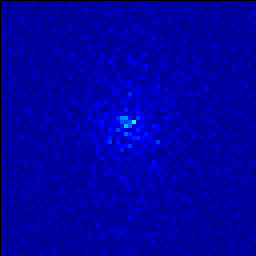}
\hspace{.75in}

(e)
\includegraphics[width=2in]{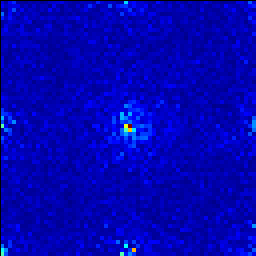}
(f)
\includegraphics[width=2in]{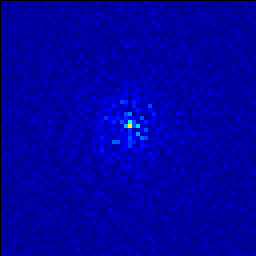}
\includegraphics[width=.75in]{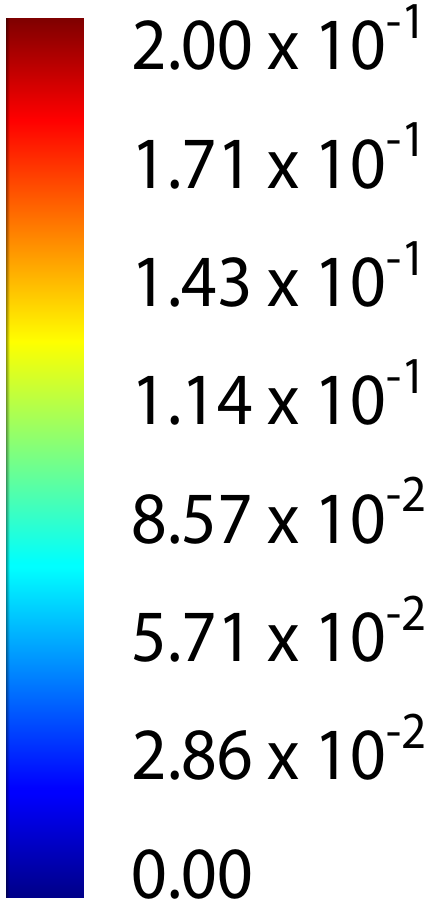}
\caption{(Color online) Magnitude of correlations:
(a) $ | \langle(\delta\hat{\rho}) (\delta\widehat{J_x}^*)\rangle | $,
(b) $ | \langle(\delta\hat{\rho}) (\delta\hat{T}^*)\rangle | $,
(c) $ | \langle(\delta\widehat{J_x}) (\delta\widehat{J_y}^*)\rangle | $,
(d) $ | \langle(\delta\widehat{J_x}) (\delta\widehat{ \rho E}^*)\rangle | $,
(e) $ | \langle(\delta\hat{\rho_1}) (\delta\hat{\rho}_4^*)\rangle | $,
(f) $ | \langle(\delta\widehat{v_x}) (\delta\hat{T}^*)\rangle | $.
In all cases the theoretical value is zero for all wave numbers.
Data range is set to $20\%$ of the normalization, as shown in the color bar.
The wave number domain represented here is
$[-3.93 \times 10^7,3.93 \times 10^7] \times [-3.93 \times 10^7,3.93 \times 10^7] $ in units
of cm$^{-1}$.}
\label{fig:correlation}
\end{figure}

We also examine correlations between different hydrodynamic variables as a function of wave number.
A set of representative correlations are presented in Figure \ref{fig:correlation}.
In each of these cases the correlation should be zero and the results show that
the normalized values are indeed quite small.
Of particular note is absence of a correlation for different components of momentum
($| \langle (\delta\widehat{ J_x}) (\delta\widehat{ J_y}^*) \rangle | \approx 0$)
validating the treatment of the stress tensor in the momentum equations.
Furthermore, the correlation between $\delta\rho_1$ and $\delta\rho_4$ is near zero,
illustrating that there is no spurious correlation
in the treatment of species diffusion.
The relaxation time for the largest wavelengths is extremely long, which is manifested
as a large statistical error at the lowest wavenumbers.

\subsection{Long-ranged Correlations in a Diffusion Barrier}
\label{DiffusionBarrierSection}

The next example considers a non-equilibrium system in which the fluctuations exhibit
long-range correlations in the presence of concentration gradients.
Here we use a hard sphere gas mixture where the three gases
(called Red, Blue, and Green or R, B, G) have equal molecular masses, taken as the
mass of argon used in the previous example.
Furthermore, Red and Blue are the same diameter, taken as that of argon, so that they are dynamically equivalent,
with the diameter of Green being a factor of 10 larger.
We set $Y_{R} = Y_{B} = 0.25$ and $Y_{G} = 0.5$ at the center of the domain and
impose gradients
of $dY_R/dy=28.935$, $dY_B/dy=90.760$ and $dY_G/dy=-119.695~\mathrm{cm}^{-1}$
for Red, Blue and Green, respectively across the domain.
These conditions produce a ``diffusion barrier'' for the Red species,
that is, the deterministic flux of Red is zero in spite of its gradient.
The initial temperature in the domain is 300K and the initial pressure is one atmosphere.
The top and bottom boundaries are no slip walls at a constant temperature of 300K
with fixed reservoir boundaries for concentrations.
Fluctuating hydrodynamics theory predicts that the spectrum of the
concentration fluctuations exhibits long-range correlations
due to the nonequilibrium conditions, even for the non-fluxing Red species, see derivation in Appendix \ref{AppendixNonEq}.

Obtaining good statistics requires a long simulation,
consequently we use a domain that is only
one cell thick in the $z$-direction, 
corresponding to an essentially two-dimensional domain. It can be shown using linearized fluctuating
hydrodynamics (i.e., small fluctuations, which corresponds to a system of thickness much larger than molecular in the $z$-direction)
that the spectrum of the concentration fluctuations is not
affected by dimensionality. Namely, upon taking a Fourier transform in the directions perpendicular to
the gradient, only the square of the perpendicular component of the wavevector enters, and in three dimensions one obtains the same
spectrum as a function of the modulus of the wavevector as one does in two dimensions. This is easily seen
in a quasi-periodic approximation, as detailed in Appendix \ref{AppendixNonEq}, but is true even in the
presence of confinement \cite{GiantFluctFiniteEffects}.

\begin{figure}
a)
\includegraphics[width=3in]{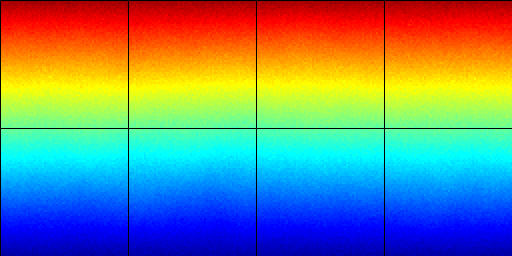}
\includegraphics[width=.6in]{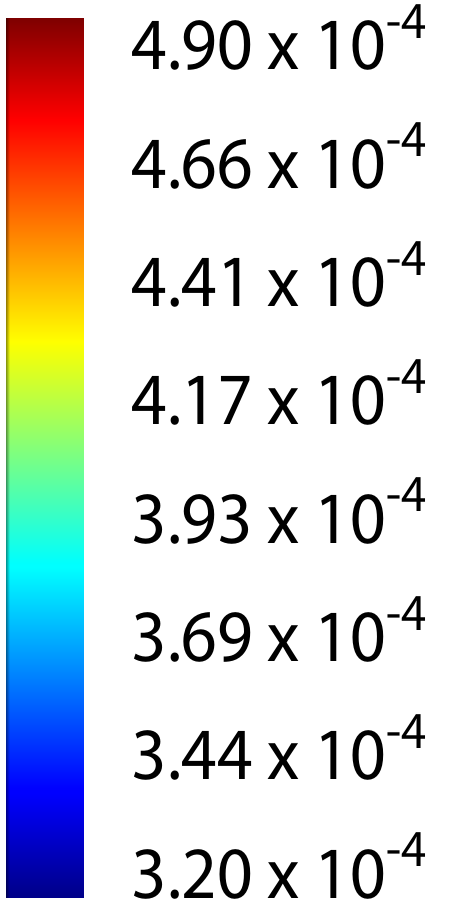}

b)
\includegraphics[width=3in]{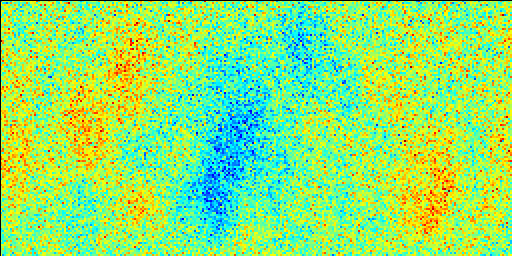}
\includegraphics[width=.6in]{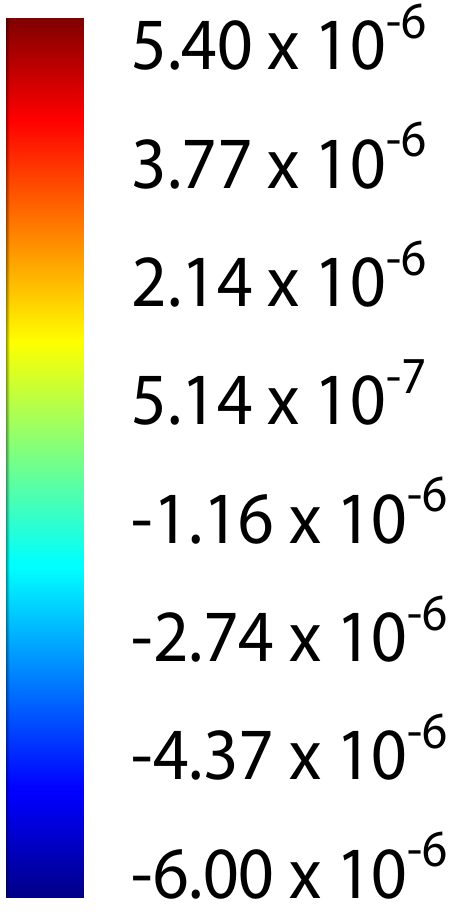}

\caption{(Color online) Typical snapshot of density of Red species in diffusion barrier simulation is shown
in (a). In (b) we subtract the background stratification and show the variation from the
background. The range of the color scale in (b) is approximately $\pm 5 \sigma_R$ where
$\sigma_R$ is the standard deviation of equilibrium fluctuations based on the value of $\rho_R$
in the center of the system.
The domain is $6.912 \times 10^{-3}$ cm $\times$ $3.456 \times 10^{-3}$ cm $\times$ $2.7\times 10^{-5}$ cm.  Units are g/cm$^3$.
}
\label{fig:giant_red}
\end{figure}

We take $\Delta x = \Delta y = \Delta z = 2.7 \times 10^{-5}$~cm on a $256 \times 128 \times 1$ domain
and a time step of $\Delta t = 10^{-10}$~s.
A no slip boundary was used in the $y$ direction and periodic
boundary conditions were used in the $x$ direction.
The simulation is run for 200000 steps to relax to a statistical steady state and
then run for an additional $2.8$ million steps, computing
$\langle(\delta\hat{\rho}_{R})(\delta\hat{\rho}_{R})^*\rangle$,
$\langle(\delta\hat{\rho}_{B})(\delta\hat{\rho}_{B})^*\rangle$ and
$\langle(\delta\hat{\rho}_{R})(\delta\hat{\rho}_{B})^*\rangle$ in Fourier space
on the vertically averaged profiles every 10 steps.
An image illustrating a typical snapshot of $\rho_{R}$ is shown in Figure \ref{fig:giant_red}a.
In Figure \ref{fig:giant_red}b we subtract off the background variation in $\rho_R$ to
more clearly show the large scale structures.
Horizontal variation of $\rho_{R}$ is apparent in the image.

\begin{figure}
\includegraphics[width=4in]{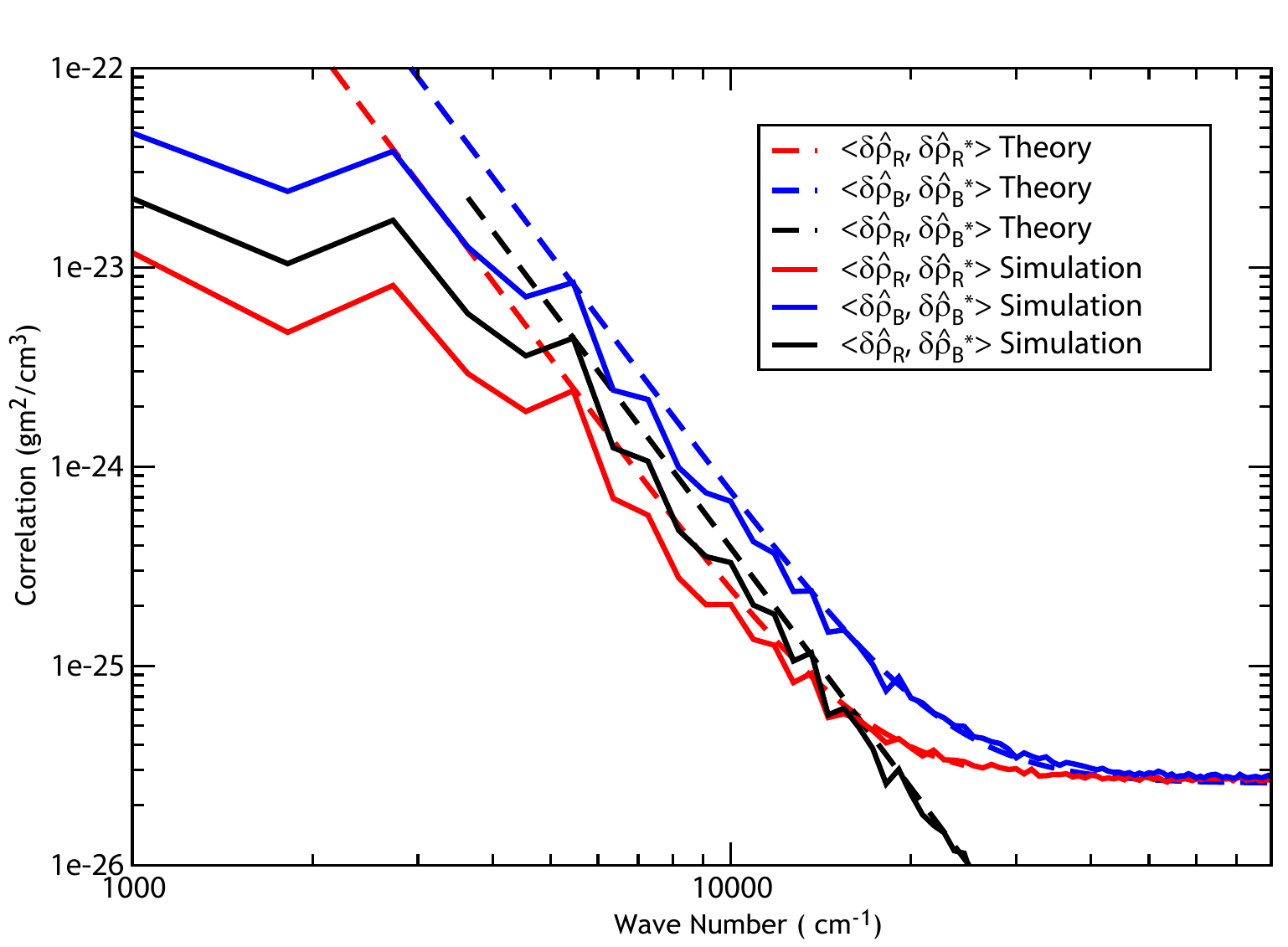}
\caption{(Color online) Static structure factor of vertically-averaged densities showing effect of giant
fluctuations.
Dashed lines represent the predictions of linearized fluctuating hydrodynamics theory, see Appendix \ref{AppendixNonEq}.  The constant limit obtained at high $k$ in the two autocorrelations corresponds to
the equilibrium values given by Eq. \ref{eq:S_c_eq_highk}.}
\label{fig:theory}
\end{figure}

To provide a quantitative characterization of the large-scale fluctuations, we plot in Figure \ref{fig:theory}
a comparison of the computed static spectra of the species densities, averaged
along the direction of the gradient, with theory (see Appendix \ref{AppendixNonEq}).
It is these spectra that can be measured experimentally using low-angle light-scattering and shadowgraph techniques.
We note that at low wave numbers the comparison breaks down because of finite size effects \cite{Zarate_07}.
Otherwise the agreement between theory and simulation is excellent.

These results show that the correlations are long-ranged with the characteristic
$k^{-4}$ power-law decay \cite{GiantFluctuations_Universal,Vailati_11,SoretDiffusion_Croccolo},
as in binary mixtures \cite{DiffusionRenormalization}, even for the
first species which has no mass flux. This demonstrates that the long-ranged
correlations are associated with the system being out of thermodynamic
equilibrium, and not with diffusive fluxes per se. Interestingly,
we find that there are giant fluctuations in all species and also
giant correlations between the fluctuations in different species.
It is anticipated that measurement of these giant fluctuations in
ternary mixtures can be used to calculate diffusion and Soret coefficients
in mixtures \cite{SoretDiffusion_Croccolo}. The main difficulty is the ability to experimentally
observe the fluctuations in different species independently.

\subsection{Diffusion-driven Rayleigh-Taylor Instability}

In this example, we illustrate how multicomponent diffusion can induce density stratification
leading to a Rayleigh-Taylor instability~\cite{BouyancyFluidsBook,MultispeciesDiffusionInstability}.
We model a four species hard sphere mixture in which two of the species
are light particles and two are heavy particles, specifically, $m_1 = m_3 < m_2 = m_4$.
For each mass, we have two different diameters, large and small, specifically,
$d_1 = d_4 < d_2 = d_3$; see Table~\ref{tab:rt}.
We initialize two layers, each of which has identical numbers of light and heavy particles in hydrostatic
equilibrium with pressure of one atmosphere at the bottom of the domain.
The result is a stably stratified isothermal initial condition of 300K with a switch in composition
in the middle of the domain as shown in Table~\ref{tab:rt}.
The simulation used a $400 \times 400 \times 200$ grid
with $\Delta x = \Delta y = \Delta z = 6 \times 10^{-6}$~cm and $\Delta t = 5 \times 10^{-12}$~s.
Gravity is set to $g = 4 \times 10^{12}~\mathrm{cm/s}^2$ in order to reduce the time
needed for the instability to develop. Boundary conditions are periodic in $x$ and $y$ with specular walls in the $z$ direction.

\begin{figure}
\includegraphics[width=2.5in]{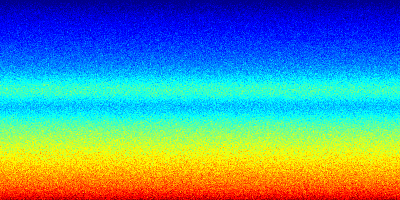}
\hspace{.2in}
\includegraphics[width=.55in]{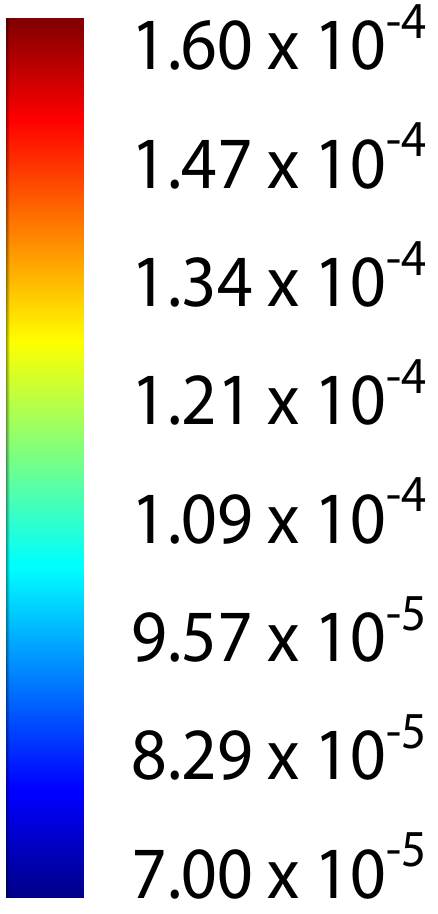}
\caption{(Color online) Vertical cross section of total density, $\rho$,
for the Rayleigh-Taylor instability simulation at
early times ($t = 2.5 \times 10^{-8}$~s).
The cross secion shown is $2.4 \times 10^{-3}$ cm $\times$ $1.2 \times 10^{-3}$ cm.  Units
are g/cm$^3$.}
\label{fig:rt_rho}
\end{figure}

\begin{figure}
(a)
\includegraphics[width=2.5in]{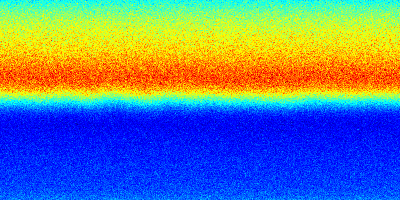}
\hspace{.2in}
\includegraphics[width=.55in]{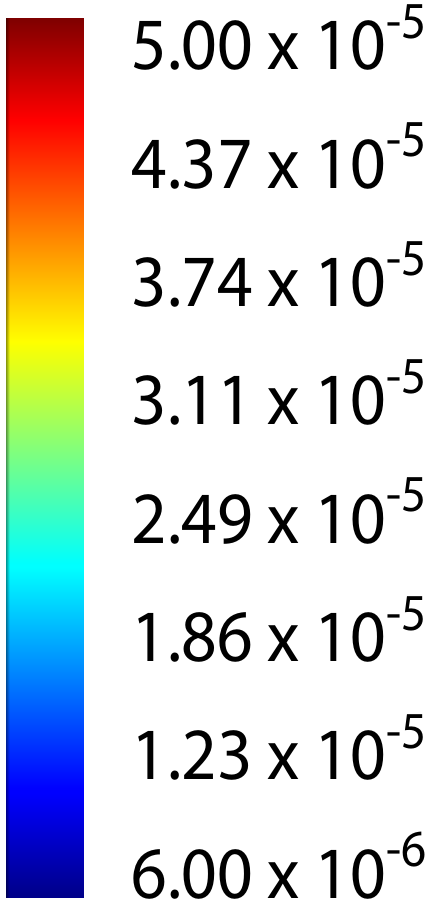}

(b)
\includegraphics[width=2.5in]{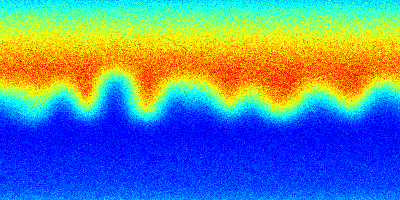}
\hspace{.2in}
\includegraphics[width=.55in]{rho2_scale.pdf}

(c)
\includegraphics[width=2.5in]{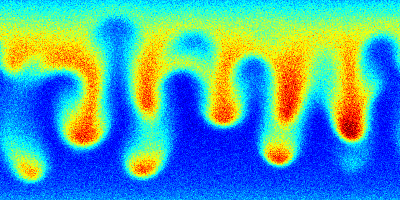}
\hspace{.2in}
\includegraphics[width=.55in]{rho2_scale.pdf}
\caption{(Color online) Vertical cross section of $\rho_2$,
the species with large, heavy particles.
Frames correspond to (a) $t = 7.5 \times 10^{-8}$~s,
(b) $t = 15.0 \times 10^{-8}$~s,
and (c) $t = 22.5 \times 10^{-8}$~s.}
The cross secion shown is $2.4 \times 10^{-3}$ cm $\times$ $1.2 \times 10^{-3}$ cm.  Units
are g/cm$^3$.
\label{fig:rt_cross}
\end{figure}

\begin{figure}
(a)
\includegraphics[width=2.5in]{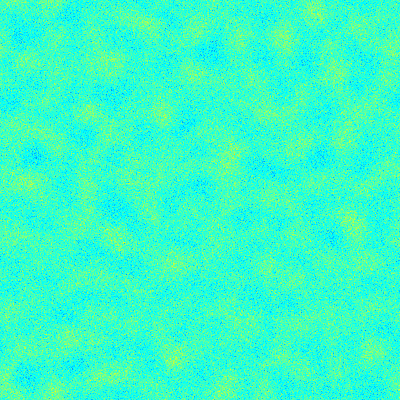}
(b)
\includegraphics[width=2.5in]{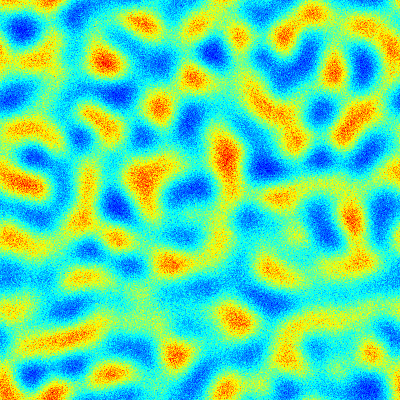}

\vspace{.1in}

(c)
\includegraphics[width=2.5in]{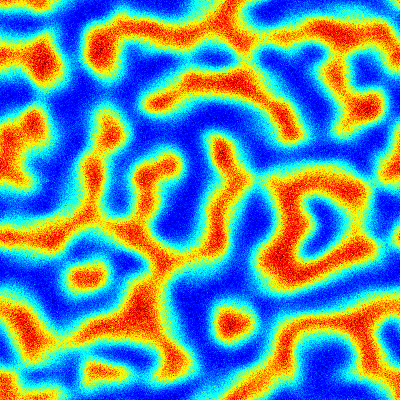}
\hspace{.2in}
\includegraphics[width=.55in]{rho2_scale.pdf}
\caption{(Color online) Horizontal slice through the center of the domain showing $\rho_2$
at (a) $t = 7.5 \times 10^{-8}$~s,
(b) $t = 15.0 \times 10^{-8}$~s,
and (c) $t = 22.5 \times 10^{-8}$~s.
The cross secion shown is $2.4 \times 10^{-3}$ cm $\times$ $2.4 \times 10^{-3}$ cm.  Units
are g/cm$^3$.}
\label{fig:rt_slice}
\end{figure}

The large particles diffuse slowly compared with the small particles  so that diffusion
of the latter dominates the early dynamics.
Initially the small, light particles are concentrated in the upper half of the domain
while the small, heavy particles are concentrated on the lower half.
This results in diffusion creating an unstable (heavier fluid on top of lighter fluid) density stratification \cite{MultispeciesDiffusionInstability},
as shown in Figure \ref{fig:rt_rho}.
At later times fluctuations within the system
trigger a Rayleigh-Taylor instability, as illustrated in
Figures \ref{fig:rt_cross} and \ref{fig:rt_slice} which show the density of species~2
(large, heavy particles).

\begin{table}
\begin{tabular}{|l|c|c|c|c|}
\hline
Species   & Molecular Weight (g) & Diameter (cm) & $Y_k$ at top & $Y_k$ at bottom \\
\hline
Species 1 & 2.0 & 2.0 $\times 10^{-8}$ & 0.4 & 0.1 \\
\hline
Species 2 & 20.0 & 20.0 $\times 10^{-8}$ & 0.4 & 0.1 \\
\hline
Species 3 &  2.0 & 20.0 $\times 10^{-8}$ & 0.1 & 0.4 \\
\hline
Species 4 &  20.0 & 2.0 $\times 10^{-8}$ & 0.1 & 0.4 \\
\hline
\end{tabular}
\caption{Molecular properties and configuration for the diffusion Rayleigh-Taylor instability.}
\label{tab:rt}
\end{table}

\subsection{Reverse Diffusion Experiment}

Our final example illustrates an application of the methodology using realistic gas properties.
In particular, we consider a three species mixture whose constituents are
molecular hydrogen, carbon dioxide, and nitrogen.
Instead of using the hard sphere model,
for this final test case the fluid properties of the gas mixture
were accurately modeled using EGLIB~\cite{EGLIB},
a general-purpose Fortran library for evaluating transport and thermodynamic
properties of gas mixtures.

This test case is qualitatively similar to the reverse diffusion
experiments of Duncan and Toor~\cite{DuncanToor}.
The domain is split into two sections (chambers in the experiment) with equal pressures, temperatures, and nitrogen densities.
The lower half of the domain is rich in carbon dioxide ($X_{H_2} = 0.1$, $X_{CO_2} = 0.4$, $X_{N_2} = 0.5$)
while the upper half is rich in hydrogen ($X_{H_2} = 0.4$, $X_{CO_2} = 0.1$, $X_{N_2} = 0.5$).
The system is initialized at $T=312.5$K and atmospheric pressure.
The simulation is performed in a $32 \times 32 \times 64$ mesh so that each half is $32^3$ with
a uniform mesh spacing of $2.7 \times 10^{-6}$~cm in each direction with periodic boundaries in $x$ and $y$ and
a specular walls in the $z$.
The simulation is run for 100000 times
steps with $\Delta t = 4. \times 10^{-13}$~s.
We note that there is no gravity in this problem and ordinary diffusion occurs for the
carbon dioxide and hydrogen.

\begin{figure}
(a)
\includegraphics[width=1in]{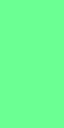}
(b)
\includegraphics[width=1in]{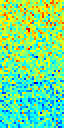}
(c)
\includegraphics[width=1in]{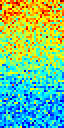}
(d)
\includegraphics[width=1in]{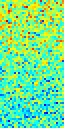}
(e)
\includegraphics[width=1in]{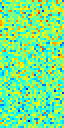}
\includegraphics[width=.55in]{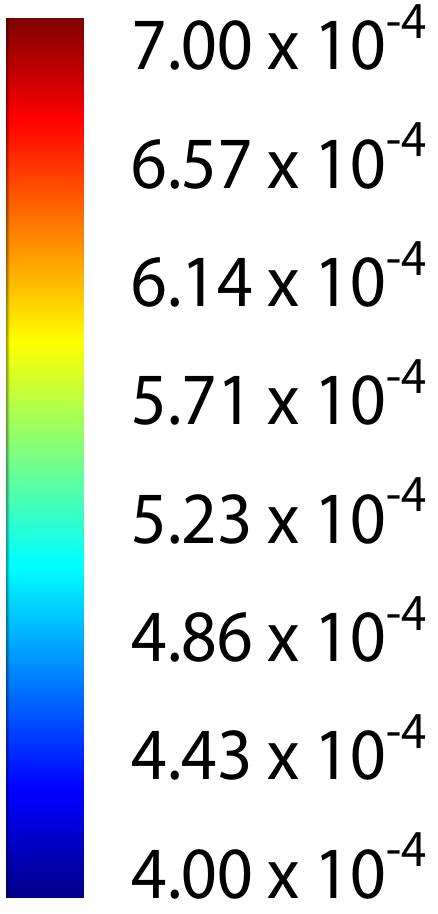}

\caption{(Color online) Slices showing temporal evolution of $\rho_{N_2}$.
Frames correspond to
(a) $t = 0.0 $~s,
(b) $t = 4.0 \times 10^{-9}$~s,
(c) $t = 8.0 \times 10^{-9}$~s,
(d) $t = 24.0 \times 10^{-9}$~s,
and (e) $t = 40.0 \times 10^{-9}$~s.
Cross section shown is $8.64 \times 10^{-5}$ cm $\times$  $2.32 \times 10^{-5}$ cm.
Units are g/cm$^3$.}
\label{fig:rev_diff}
\end{figure}

In Figure \ref{fig:rev_diff}, we show slices of nitrogen density, $\rho_{N_2}$, at a sequence of times.
At $t = 4.0 \times 10^{-9}$~s we see that, although its concentration is initially uniform,
due to interdiffusion effects there has been a flux of $N_2$ into the upper half of the domain.
In spite of the adverse gradient, diffusive transport continues to increase the amount of $N_2$ in
the upper half as seen in Figure \ref{fig:rev_diff}c.
This ``reverse diffusion'' of nitrogen is driven by the rapid diffusion of $H_2$ out of the upper
half, as compared with the slow diffusion of carbon dioxide into it.
The last two frames in Figure \ref{fig:rev_diff} show the slow
return towards uniform nitrogen concentration.

\begin{figure}
\includegraphics[width=3in]{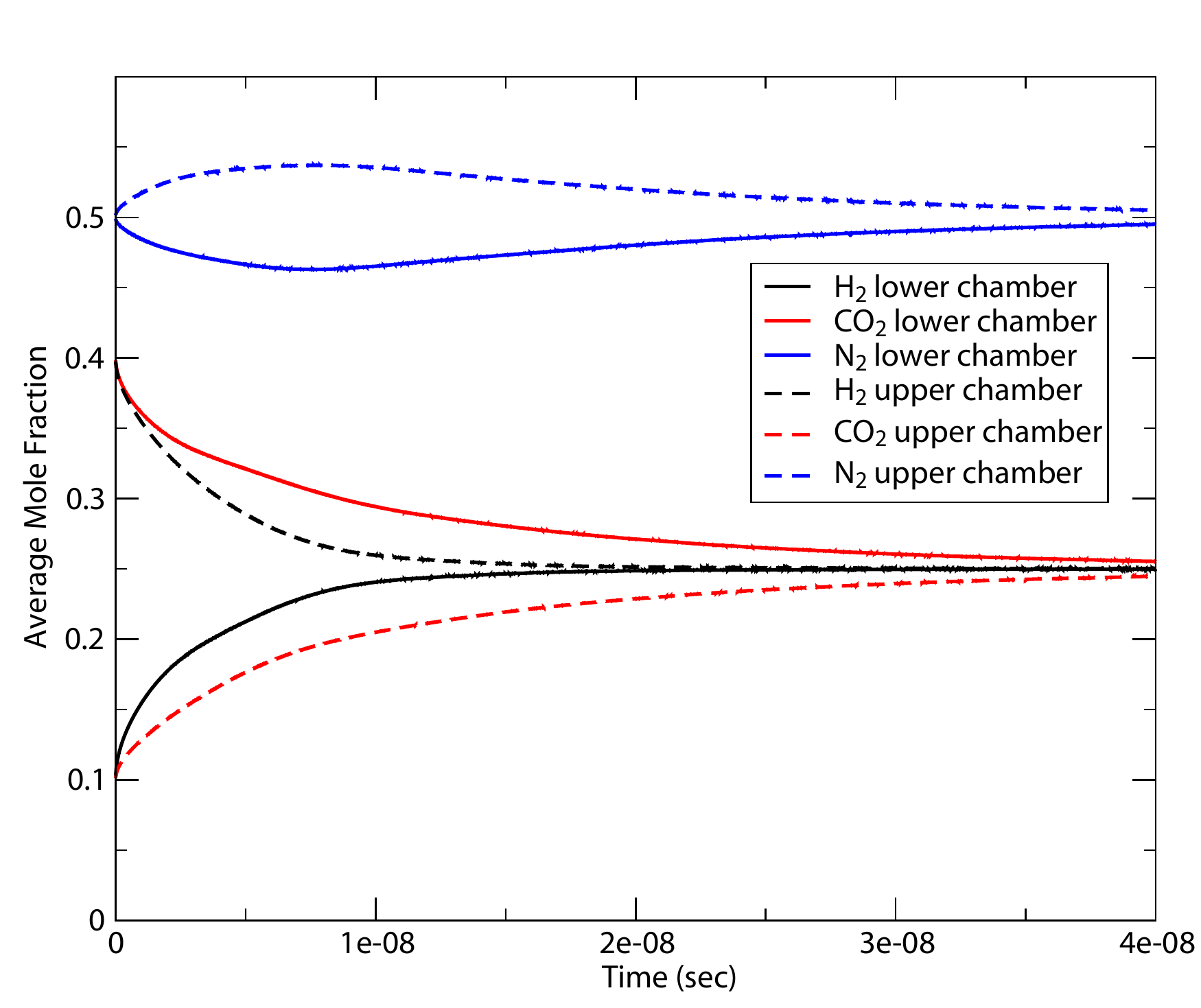}
\caption{(Color online) Time evolution of average composition in upper and lower halves of the domain.}
\label{fig:rev_hist}
\end{figure}

This reverse diffusion phenomenon is shown quantitatively in Figure \ref{fig:rev_hist}
where we plot the time history of the average mole fractions in the upper and lower halves of the domain.
We see that there is a flux of $N_2$ into the upper half until approximately
$t = 8.0 \times 10^{-9}$~s in spite of the adverse $N_2$ gradient, at which point a diffusion barrier occurs (gradient of $N_2$ without a flux).
At later times the diffusion is ``normal'' for all three species.
We note that the data in Figure \ref{fig:rev_hist} is in qualitative agreement with
the experimental results of Duncan and Toor; the primary differences between the simulation and
the experimental set-up are the size and geometry of the system.
Thermal fluctuations do not play a significant
role in this numerical experiment, although the appearance of giant fluctuations due to the transient concentration gradients is expected.
The primary purpose of this final numerical test is to demonstrate the ability of
our implementation to simulate gas mixtures using transport and thermodynamic
properties given by the EGLIB library, allowing quantitative comparison with experiments.

\section{Conclusions and Future Work}
\label{sec:conclusions}

The four examples in the previous section confirm the accuracy of our
numerical formulation for the multispecies fluctuating Navier-Stokes equations.
Furthermore, they illustrate some of the interesting phenomena unique to such fluid mixtures.
While these numerical examples were all gas mixtures the methodology is directly
extendable to liquids, the main challenge being the formulation of
accurate thermodynamic and transport properties \cite{TernaryEquilibriumFluct}. Work in this direction is currently
underway and fluctuating hydrodynamics should prove useful for experimental studies
of the properties of liquid mixtures \cite{SoretDiffusion_Croccolo}.

A numerical limitation of the methodology presented here is that the stochastic PDE solver
is explicit. This restricts practical application of the method to the study of
phenomena of mesoscopic duration ($\approx$ microsecond) given the magnitude
of the algorithm's stable time step. This restriction is particularly
severe for liquid mixtures for which there are orders of magnitude of separation between the
fast acoustic, intermediate viscous, and slow diffusive time scales, at which phenomena of
interest occur (e.g., minutes or hours for giant fluctuation experiments \cite{Vailati_11}).
To lift the time step limitation we are investigating low-Mach number approximations
for mixtures \cite{DonevLowMach2013}.
Another important avenue of research we are presently persuing is to extend our
formulation to non-ideal multispecies mixtures of non-ideal fluids \cite{TernaryEquilibriumFluct,NewIrrevThermoBook}.

We are also extending the formulation to reacting, multi-component mixtures,
which will lay the groundwork for the investigation of a
wide variety phenomena combining hydrodynamic and chemical fluctuations.
Numerical methods for fluctuating reaction-diffusion systems date back to the
early 1970's~\cite{Gillespie1976403,NicolisPrigogine,GardinerBook}
but these methods neglect all hydrodynamic transport other than diffusion
and typically diffusion is also simplified~\cite{GillespieReview2013}.
While this is a good approximation
for many phenomena, a complete description of transport
is required for combustion.


\begin{acknowledgments}
The work at LBNL was supported by the Applied Mathematics Program of the U.S. DOE Office
of Advance Scientifc Computing Research under contract DE-AC02005CH11231.
A. Donev was supported in part by the National Science Foundation under grant DMS-1115341
and the Office of Science of the U.S. Department of Energy through Early Career award number DE-SC0008271.
This research used resources of the National Energy Research Scientific Computing Center,
which is supported by the Office of Science of
the U.S. Department of Energy under Contract No. DE-AC02-05CH11231.
\end{acknowledgments}


\appendix

\section{Variances in a Multicomponent Gas Mixture\label{AppendixEq}}

For species $i$ the number of molecules in a volume $V$ is $N_i = V \rho_i / m_i$ where $m_i$ is the mass of a molecule.
At equilibrium for an ideal gas this number is Poisson distributed and independent of other species so $\langle \delta N_i \delta N_j \rangle = \bar{N}_i \delta_{ij}$ and,
\begin{equation}
\langle \delta \rho_i \delta \rho_j \rangle_{eq}
= \frac{\bar{\rho}_i^2}{\bar{N}_i} \delta_{ij}
= \frac{\bar{\rho}_i^2 k_B \bar{T}}{\bar{P}\bar{X}_i V} \delta_{ij},
\end{equation}
where the subscript ``eq'' indicates an equilibrium result.
At equilibrium the variance of density is thus,
\begin{equation}
\langle \delta \rho^2 \rangle_\mathrm{eq} =
\sum_i \sum_j \langle \delta \rho_i \delta \rho_j \rangle_\mathrm{eq}
= \sum_i \frac{\bar{\rho}_i^2}{\bar{N}_i}
= \frac{\bar{\rho}^2}{\bar{N}}\sum_i \frac{Y_i^2}{X_i}
= \frac{\bar{\rho}^2 k_B \bar{T}}{\bar{P}V}\sum_i \frac{Y_i^2}{X_i}.
\end{equation}
From this,
\begin{equation}
\langle \delta \rho^2 \rangle_\mathrm{eq}
= \zeta \langle \delta \rho^2 \rangle_\mathrm{eq}^{(1)}
\qquad\mathrm{where}\qquad
\zeta = \sum_i \frac{Y_i^2}{X_i}
\end{equation}
and $\langle \delta \rho^2 \rangle_\mathrm{eq}^{(1)} = \bar{\rho}^2/\bar{N}$ is the variance for a single species gas
at the same density, temperature, and pressure (i.e., same $\bar{N} = \bar{P}V/k_B \bar{T}$).
Note that all of the expressions in this appendix may be generalized easily to spatial correlations, for example,
\begin{equation}
\langle \delta \rho(\mathbf{r}) \delta \rho(\mathbf{r}')\rangle
= \frac{\bar{\rho}^2 k_B \bar{T}}{\bar{P}} \delta(\mathbf{r} - \mathbf{r}')
\sum_i \frac{Y_i^2}{X_i}.
\end{equation}

For the variance and correlations of momentum the results are the same as
those for a single species gas when $\bar{\mathbf{v}} = 0$, specifically,
\begin{eqnarray}
\langle \delta \rho \delta \mathbf{J} \rangle_\mathrm{eq} &=& 0 \\
\langle \delta J_{\alpha} \delta J_{\beta} \rangle_\mathrm{eq} &=&
\frac{\bar{\rho} k_B \bar{T}}{V} \delta_{\alpha\beta}.
\end{eqnarray}
Similarly, for velocity,
\begin{eqnarray}
\langle \delta \rho \delta \mathbf{v} \rangle_\mathrm{eq} &=& 0 \\
\langle \delta v_{\alpha} \delta v_{\beta} \rangle_\mathrm{eq} &=&
\frac{k_B \bar{T}}{\bar{\rho} V} \delta_{\alpha\beta}.
\end{eqnarray}

Finally, for energy fluctuations,
\begin{eqnarray}
\delta \mathcal{E} = \delta (\rho E)
&=& \delta \left( \frac{1}{2} \rho v^2+ \sum_i \rho_i e_i(T) \right) \\
&=& \bar{\rho} \bar{\mathbf{v}} \cdot \delta\mathbf{v}
+ \frac{1}{2} \bar{v}^2 \delta \rho
+ \sum_i \delta \rho_i e_i(\bar{T})
+ \bar{\rho} c_{v}(\bar{T})\, \delta T,
\end{eqnarray}
where
\begin{equation}
c_v(T) = \frac{1}{\rho} \sum_i \rho_i c_{v,i}(T) = \sum_i Y_i c_{v,i}(T)
\end{equation}
is the molar averaged specific heat.
%
%
Note that if $c_v$ is independent of temperature then $e_i(\bar{T}) = c_{v,i} \bar{T}$.

The resulting variance and correlations for energy are,
\begin{eqnarray}
\langle \delta \rho \delta \mathcal{E} \rangle_\mathrm{eq} &=&
\sum_i e_i(\bar{T}) \langle \delta \rho_i^2 \rangle_\mathrm{eq}; \\
\langle \delta \mathbf{J} \delta \mathcal{E} \rangle_\mathrm{eq} &=& 0; \\
\langle \delta \mathcal{E}^2 \rangle_\mathrm{eq} &=&
\sum_i e_i(\bar{T})^2 \langle \delta \rho_i^2 \rangle_\mathrm{eq}
+ \bar{\rho}^2 c_{v}(\bar{T})^2 \langle \delta T^2 \rangle_\mathrm{eq}.
\end{eqnarray}
Similarly for temperature,
\begin{eqnarray}
\langle \delta \rho \delta T \rangle_\mathrm{eq} &=& 0; \\
\langle \delta \mathbf{v} \delta T \rangle_\mathrm{eq} &=& 0; \\
\langle \delta T^2 \rangle_\mathrm{eq} &=& \frac{k_B T^2}{\bar{\rho} c_v(\bar{T}) V}.
\end{eqnarray}

\section{Giant fluctuations in a ternary mixture\label{AppendixNonEq}}

This appendix outlines the fluctuating hydrodynamics theory for
the long-range correlations of concentration fluctuations in a ternary mixture,
in order to model the simulations described in Section \ref{DiffusionBarrierSection}.
We neglect the Dufour effect and assume the system to be isothermal,
taking contributions from temperature fluctuations to be of higher order.
Furthermore, we neglect gravity, assume the system is incompressible,
and take the density and transport coefficients to be constant.
We consider a ``bulk'' system \cite{Zarate_07},
i.e., we neglect the influence of the boundaries.
This gives an accurate approximation for wavenumbers that are large compared
to the inverse height of the domain; for smaller wavenumbers the
boundaries are expected to suppress the giant fluctuations~\cite{Zarate_07,GiantFluctFiniteEffects,Vailati_11}.
Lastly, we initially ignore the equilibrium fluctuations in the calculation
and simply add them to the non-equilibrium contribution at the end.
This is not necessary and the additional stochastic diffusive fluxes can easily be accounted for
at the expense of algebraic complications. This confirms that the equilibrium fluctuations
enter additively to the nonequilibrium ones calculated here, as confirmed by an anonymous reviewer.

As in the problem considered in Section~\ref{DiffusionBarrierSection},
we assume all of the concentration gradients are in the same
direction (say, the $y$ axis). The incompressibility constraint
is most easily handled by applying a $\grad\times\grad\times$
operator to the momentum equation to obtain a system involving only
the component of the velocity parallel to the gradient
(in this case, the $y$-direction).~\cite{Zarate_07}
With the above, the momentum and concentration equations yield,
\begin{eqnarray*}
\partial_{t}\left(\grad^{2}v_{\parallel}\right) &=& \nu\grad^{2}\left(\grad^{2}v_{\parallel}\right)+
\rho_{0}^{-1}\grad\times\grad\times\left(\grad\cdot\widetilde{\M{\Pi}}\right)
\label{eq:LLNS_comp_v_simp-2}\\
\partial_{t}\left(\d{\mathbf{Y}}\right) &=& -v_{\parallel} \V{f}
+\M{\DonevDiffusion}\grad^{2}\left(\d{\mathbf{Y}}\right),\label{eq:LLNS_comp_c_simp-2}
\end{eqnarray*}
where $\rho_0$ is the constant density,
and $\nu = \eta/\rho_0$ is the kinematic viscosity.
Here $\M{\DonevDiffusion}=\rho^{-1}\M C\left(\frac{\partial\V X}{\partial\V Y}\right)$
is a matrix of diffusion coefficients and $\partial\V X/\partial\V Y$
is the Jacobian of the transformation from mass to mole fractions,
which is a function of the mean molecular mass and the individual species molecular masses.
Here $\V{f}=d\langle{\mathbf{Y}}\rangle/dy$ are the imposed mass fraction gradients
and $\d{\mathbf{Y}}=\mathbf{Y}-\langle{\mathbf{Y}}\rangle$ is the concentration fluctuation.

This system of equations can be most easily solved in the Fourier
domain, where it becomes
\begin{eqnarray}
\partial_{t}\hat{v}_{\parallel} &=&
-\nu k^{2}\hat{v}_{\parallel}+\hat{F}
\label{eq:LLNS_comp_v_f}\\
\partial_{t}\left(\d{\hat{\mathbf{Y}}}\right) &=&
-\hat{v}_{\parallel} \V{f}
-k^{2}\M{\DonevDiffusion}\d{\hat{\mathbf{Y}}},
\label{eq:LLNS_comp_c_f}
\end{eqnarray}
and the covariance of the random forcing $\hat{F}$ is (see (5.12)
in Ref. \cite{Zarate_07})
\[
\av{\hat{F}\hat{F}^{\star}}=\frac{2k_{B}T_{0}}{\rho_{0}}\nu k_{\perp}^{2},
\]
where $\V k_{\perp}$ is the component of the wavevector in the plane
perpendicular to the gradient and $T_0$ is the constant temperature.
The equilibrium covariance of the fluctuations,
written as a matrix of static structure factors,
\[
\M S=\left[\begin{array}{cc}
\av{\hat{v}_{\parallel}\hat{v}_{\parallel}^{\star}} & \av{\left(\d{\hat{\mathbf{Y}}}\right)\hat{v}_{\parallel}^{\star}}\\
\av{\hat{v}_{\parallel}\left(\d{\hat{\mathbf{Y}}}\right)^{\star}} & \av{\left(\d{\hat{\mathbf{Y}}}\right)\left(\d{\hat{\mathbf{Y}}}\right)^{\star}}
\end{array}\right]
\]
can be obtained most directly by writing the equations (\ref{eq:LLNS_comp_v_f},\ref{eq:LLNS_comp_c_f})
in the form of an Ornstein-Uhlenbeck (OU) process,
\[
\partial_{t}\left[\begin{array}{c}
\hat{v}_{\parallel}\\
\d{\hat{\mathbf{Y}}}
\end{array}\right]=\left[\begin{array}{cc}
-\nu k^{2} & 0\\
-\V{f} & -k^{2}\M{\DonevDiffusion}
\end{array}\right]\left[\begin{array}{c}
\hat{v}_{\parallel}\\
\d{\hat{\mathbf{Y}}}
\end{array}\right]+\left[\begin{array}{c}
\hat{F}\\
\V 0
\end{array}\right]=\M M\left[\begin{array}{c}
\hat{v}_{\parallel}\\
\d{\hat{\mathbf{Y}}}
\end{array}\right]+\V m,
\]
and using the well-known equation for the equilibrium covariance of
an OU process \cite{GardinerBook} (see, for example, derivation in
Ref. \cite{Donev_10}),
\begin{equation}
\M M\M S+\M S\M M^{\star}=-\av{\V m\V m^{\star}}.\label{eq:S_c_eq}
\end{equation}
This is a linear system of equations for the static structure factors
that can easily be solved using computer algebra systems. Note that here
the equation for the third species is redundant and it is simpler to develop
the theory by considering the equations for only the first two species.

Turning now to the specific example of a ternary mixture considered
in Section~\ref{DiffusionBarrierSection}: The molecular masses are identical and therefore
mole and mass fractions are the same, $\partial\V X/\partial\mathbf{Y}=\M I$.
Furthermore, the first two of the three species are indistinguishable
so $\M{\DonevDiffusion}$ has the simple form,
\[
\M{\DonevDiffusion}=\left[\begin{array}{cc}
\DonevDiffusion_{1} & \DonevDiffusion_{2}\\
\DonevDiffusion_{2} & \DonevDiffusion_{1}
\end{array}\right].
\]
The system is set up with a diffusion barrier, that is,
there is no deterministic flux for the first species. This implies that,
\[
f_1 = \frac{d\langle{Y}_1\rangle}{dy} = -\frac{\DonevDiffusion_{2}}{\DonevDiffusion_{1}} \frac{d\langle{Y}_2\rangle}{dy}.
\]
We consider the spectrum of the fluctuations of the partial
densities averaged along the direction of the gradient, as is measured
in experiments \cite{Vailati_11,GiantFluctuations_Universal,SoretDiffusion_Croccolo},
i.e., we take $k_{\parallel}=0$, $\V k=\V k_{\perp}$. The solution
of (\ref{eq:S_c_eq}) gives the non-equilibrium contribution to the
static structure factor for vertically-averaged concentration fluctuations
to be
\[
\av{\left(\d{\hat{\mathbf{Y}}}\right)\left(\d{\hat{\mathbf{Y}}}\right)^{\star}}_{\text{neq}}=K\left[\begin{array}{cc}
{\frac{8\,{\DonevDiffusion_{{1}}}^{3}+5\,{\DonevDiffusion_{{1}}}^{2}\nu-\nu\,{\DonevDiffusion_{{2}}}^{2}}{{\DonevDiffusion_{{1}}}^{2}\DonevDiffusion_{{2}}}} & -{\frac{3\,{\DonevDiffusion_{{1}}}^{2}\nu+\nu\,{\DonevDiffusion_{{2}}}^{2}+4\,{\DonevDiffusion_{{1}}}^{3}+4\,{\DonevDiffusion_{{2}}}^{2}\DonevDiffusion_{{1}}}{{\DonevDiffusion_{{2}}}^{2}\DonevDiffusion_{{1}}}}\\
\noalign{\medskip}-{\frac{3\,{\DonevDiffusion_{{1}}}^{2}\nu+\nu\,{\DonevDiffusion_{{2}}}^{2}+4\,{\DonevDiffusion_{{1}}}^{3}+4\,{\DonevDiffusion_{{2}}}^{2}\DonevDiffusion_{{1}}}{{\DonevDiffusion_{{2}}}^{2}\DonevDiffusion_{{1}}}} & {\frac{{\DonevDiffusion_{{2}}}^{4}\nu+2\,{\DonevDiffusion_{{1}}}^{4}\nu+2\,{\DonevDiffusion_{{1}}}^{5}+{\DonevDiffusion_{{1}}}^{2}\nu\,{\DonevDiffusion_{{2}}}^{2}+2\,{\DonevDiffusion_{{2}}}^{4}\DonevDiffusion_{{1}}+4\,{\DonevDiffusion_{{1}}}^{3}{\DonevDiffusion_{{2}}}^{2}}{{\DonevDiffusion_{{1}}}^{2}{\DonevDiffusion_{{2}}}^{3}}}
\end{array}\right]
\]
where the common pre-factor is
\[
K={\frac{{\it k_{B}T}}{2\rho\,{k}^{4}}}\cdot\frac{\DonevDiffusion_{{1}}\DonevDiffusion_{{2}}}{\left(\DonevDiffusion_{{1}}-\DonevDiffusion_{{2}}\right)\left(\DonevDiffusion_{{1}}+\DonevDiffusion_{{2}}\right)\left(\DonevDiffusion_{{1}}+\nu+\DonevDiffusion_{{2}}\right)\left(\DonevDiffusion_{{1}}+\nu-\DonevDiffusion_{{2}}\right)}
\cdot f_1^2.
\]
The equilibrium static structure factor for the mixture of ideal
gases considered here is
\begin{equation}
\av{\left(\d{\hat{\mathbf{Y}}}\right)\left(\d{\hat{\mathbf{Y}}}\right)^{\star}}_{\text{eq}}
=\rho_{0}^{-2}\left[\begin{array}{cc}
m_{1}\langle{\rho}_{1}\rangle & 0\\
0 & m_{2}\langle{\rho}_{2}\rangle
\end{array}\right] \;\; ,
\label{eq:S_c_eq_highk}
\end{equation}
which is to be added to the nonequilibrium contribution to obtain the full spectrum, as shown in Fig. \ref{fig:theory}.

In the case of liquids, the Schmidt number is very large and
$\DonevDiffusion_{1}\ll\nu$ and $\DonevDiffusion_{2}\ll\nu$ and the expressions simplify considerably,
\[
\av{\left(\d{\hat{\mathbf{Y}}}\right)\left(\d{\hat{\mathbf{Y}}}\right)^{\star}}_{\text{neq}}=
{\frac{\DonevDiffusion_{{1}}\DonevDiffusion_{{2}}
{\it k_{B}T {|\nabla_z Y_1|}^{2} }}{2\eta\,{k}^{4}
\left(\DonevDiffusion_{{1}}-\DonevDiffusion_{{2}}\right)\left(\DonevDiffusion_{{1}}+\DonevDiffusion_{{2}}\right)}}
\left[\begin{array}{cc}
{\frac{5\,{\DonevDiffusion_{{1}}}^{2}-{\DonevDiffusion_{{2}}}^{2}}{{\DonevDiffusion_{{1}}}^{2}\DonevDiffusion_{{2}}}} & -{\frac{3\,{\DonevDiffusion_{{1}}}^{2}+{\DonevDiffusion_{{2}}}^{2}}{{\DonevDiffusion_{{2}}}^{2}\DonevDiffusion_{{1}}}}\\
\noalign{\medskip}-{\frac{3\,{\DonevDiffusion_{{1}}}^{2}+{\DonevDiffusion_{{2}}}^{2}}{{\DonevDiffusion_{{2}}}^{2}\DonevDiffusion_{{1}}}} & {\frac{{\DonevDiffusion_{{2}}}^{4}+2\,{\DonevDiffusion_{{1}}}^{4}+{\DonevDiffusion_{{1}}}^{2}{\DonevDiffusion_{{2}}}^{2}}{{\DonevDiffusion_{{1}}}^{2}{\DonevDiffusion_{{2}}}^{3}}}
\end{array}\right].
\]
This can be more straightforwardly obtained by simply deleting the inertial term $\partial_{t}\hat{v}_{\parallel}$
on the left-hand side of the momentum equation (\ref{eq:LLNS_comp_v_f}) \cite{GiantFluctFiniteEffects}.
Note however that the Schmidt number is not large for gas mixtures and one must retain the complete expression
to obtain a good match to the numerical results.

\end{document}